\documentclass{article}

\PassOptionsToPackage{numbers}{natbib}

\usepackage[preprint]{neurips_2025}

\usepackage[utf8]{inputenc}
\usepackage[T1]{fontenc}
\usepackage{hyperref}
\usepackage{url}
\usepackage{booktabs}
\usepackage{amsfonts}
\usepackage{nicefrac}
\usepackage{microtype}
\usepackage{xcolor}
\usepackage{rotating}
\usepackage{afterpage}

\usepackage{multirow}
\usepackage{tabularx}
\usepackage{array}
\usepackage{pifont}
\usepackage{graphicx}
\usepackage{tcolorbox}

\makeatletter
\renewcommand{\@makefnmark}{\hbox{\@textsuperscript{\normalfont\@thefnmark}}}
\renewcommand{\thefootnote}{\arabic{footnote}}
\makeatother

\newcolumntype{L}[1]{>{\raggedright\arraybackslash}p{#1}}

\newtcolorbox{takeawaybox}{
  colback=black!3,
  colframe=black!40,
  title=\textbf{Key Takeaways},
  fonttitle=\normalsize\bfseries,
  boxrule=0.6pt,
  arc=1.5mm,
  left=5pt,
  right=5pt,
  top=5pt,
  bottom=5pt
}

\title{
\fontsize{16pt}{19pt}\selectfont
GenAI for Systems:\\Recurring Challenges and Design Principles from Software to Silicon
}
\author{%
  \textbf{Arya Tschand}\textsuperscript{*$\dagger$}, 
  \textbf{Chenyu Wang}\textsuperscript{$\dagger$}, 
  \textbf{Zishen Wan}\textsuperscript{$\dagger$}, 
  \textbf{Andrew Cheng}\textsuperscript{$\ddagger$}, 
  \textbf{Ioana Cristescu}\textsuperscript{$\ddagger$}, \\
  \textbf{Kevin He}\textsuperscript{$\ddagger$}, 
  \textbf{Howard Huang}\textsuperscript{$\ddagger$}, 
  \textbf{Alexander Ingare}\textsuperscript{$\ddagger$}, 
  \textbf{Akseli Kangaslahti}\textsuperscript{$\ddagger$}, 
  \textbf{Sara Kangaslahti}\textsuperscript{$\ddagger$}, \\
  \textbf{Theo Lebryk}\textsuperscript{$\ddagger$}, 
  \textbf{Hongjin Lin}\textsuperscript{$\ddagger$}, 
  \textbf{Jeffrey Jian Ma}\textsuperscript{$\ddagger$}, 
  \textbf{Alexandru Meterez}\textsuperscript{$\ddagger$}, 
  \textbf{Clara Mohri}\textsuperscript{$\ddagger$}, \\
  \textbf{Depen Morwani}\textsuperscript{$\ddagger$}, 
  \textbf{Sunny Qin}\textsuperscript{$\ddagger$}, 
  \textbf{Roy Rinberg}\textsuperscript{$\ddagger$}, 
  \textbf{Paula Rodriguez-Diaz}\textsuperscript{$\ddagger$}, 
  \textbf{Alyssa Mia Taliotis}\textsuperscript{$\ddagger$}, \\
  \textbf{Pernille Undrum Fathi}\textsuperscript{$\ddagger$}, 
  \textbf{Rosie Zhao}\textsuperscript{$\ddagger$}, 
  \textbf{Todd Zhou}\textsuperscript{$\ddagger$}, 
  \textbf{Vijay Janapa Reddi}
}

\begin{document}

% Adjust page geometry for wider text
\newgeometry{
  textheight=9in,
  textwidth=6.5in,
  top=1in,
  headheight=12pt,
  headsep=25pt,
  footskip=30pt
}

\maketitle

% Affiliation displayed below author list
\vspace{-0.9cm}
\begin{center}
{\large Harvard University}
\end{center}

\vspace{0.2cm}

% Author role footnotes
\renewcommand{\thefootnote}{*}
\footnotetext{Corresponding author: \texttt{aryatschand@g.harvard.edu}}
\renewcommand{\thefootnote}{$\dagger$}
\footnotetext{Equal contribution. These authors led the coordination and synthesis of this paper.}
\renewcommand{\thefootnote}{$\ddagger$}
\footnotetext{Listed alphabetically; these authors contributed as part of the \href{https://harvard-edge.github.io/cs249r_fall2025/}{CS249r: Architecture 2.0, Agentic AI for Computer Systems Design} graduate seminar at Harvard.}

%%
%% Abstract
%%
\begin{abstract}
Generative AI is reshaping how computing systems are designed, optimized, and built, yet research remains fragmented across software, architecture, and chip design communities. This paper takes a cross-stack perspective, examining how generative models are being applied from code generation and distributed runtimes through hardware design space exploration to RTL synthesis, physical layout, and verification. Rather than reviewing each layer in isolation, we analyze how the same structural difficulties and effective responses recur across the stack. Our central finding is one of convergence. Despite the diversity of domains and tools, the field keeps encountering five recurring challenges (the feedback loop crisis, the tacit knowledge problem, trust and validation, co-design across boundaries, and the shift from determinism to dynamism) and keeps arriving at five design principles that independently emerge as effective responses (embracing hybrid approaches, designing for continuous feedback, separating concerns by role, matching methods to problem structure, and building on decades of systems knowledge). We organize these into a challenge--principle map that serves as a diagnostic and design aid, showing which principles have proven effective for which challenges across layers. Through concrete cross-stack examples, we show how systems navigate this map as they mature, and argue that the field needs shared engineering methodology, including common vocabularies, cross-layer benchmarks, and systematic design practices, so that progress compounds across communities rather than being rediscovered in each one. Our analysis covers more than 275 papers spanning eleven application areas across three layers of the computing stack, and distills open research questions that become visible only from a cross-layer vantage point.
\end{abstract}

% \begin{figure}[h]
%     \centering
%     \includegraphics[width=\textwidth]{Figures/CS249R_Survey_Paper_Figure.drawio.png}
%     \caption{Overview of sections}
%     \label{fig:paper_sections}
% \end{figure}

\section{Introduction}
\label{sec:intro}

\begin{figure}[h]
    \centering
    \includegraphics[width=0.95\textwidth]{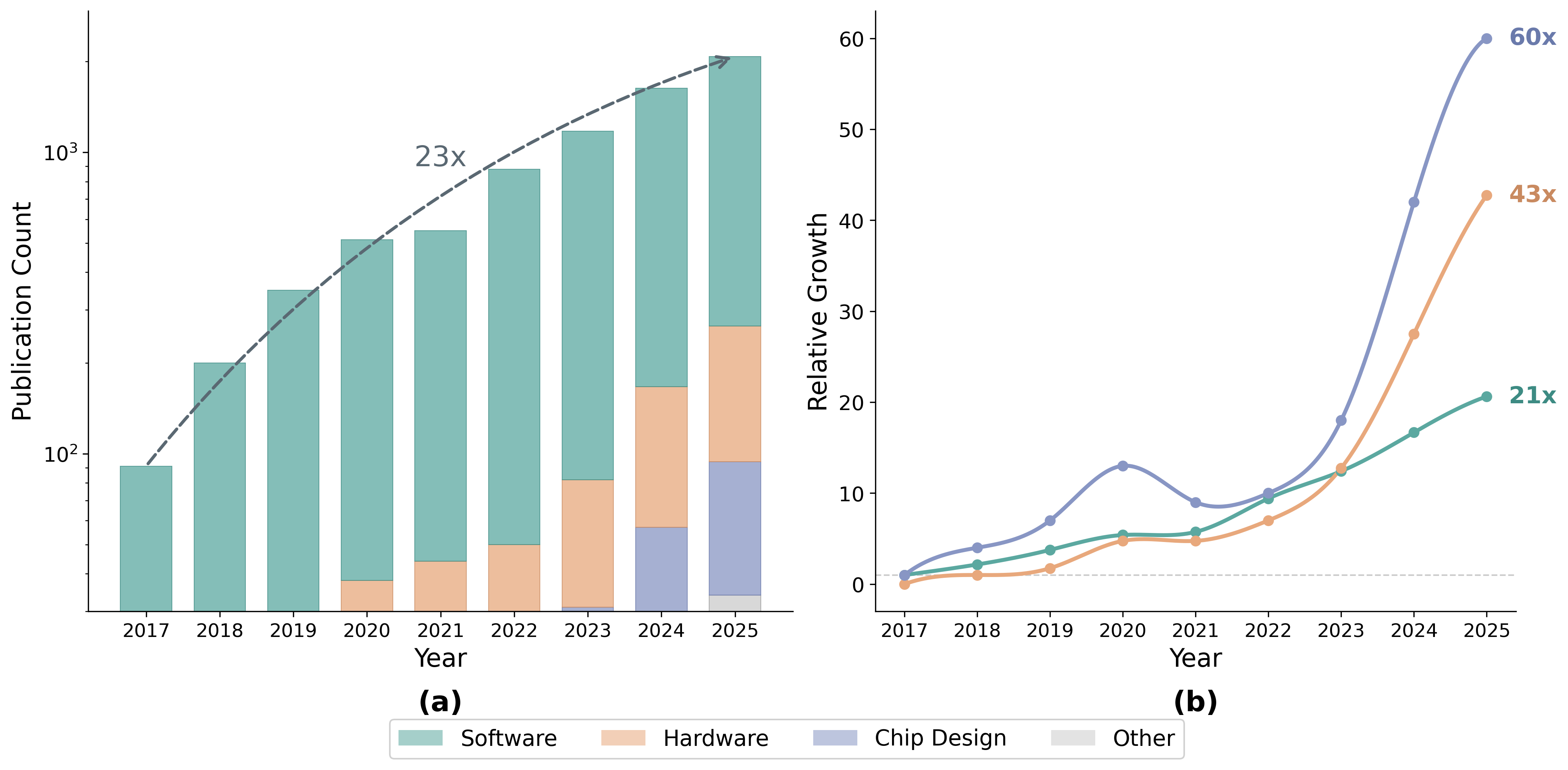}
    \caption{Growth of AI for Systems publications on arXiv, based on an analysis of over 7,800 papers (2017--2025). (a)~Publication count by domain (log scale), showing a 23$\times$ overall increase. (b)~Per-domain growth normalized to 2017 levels: although Software dominates in volume, Hardware (43$\times$) and Chip Design (60$\times$) are rapidly catching up with significantly higher growth rates compared to Software (21$\times$), reflecting the field's broadening from software-centric applications toward the full computing stack.}
    \label{fig:ml_systems_growth_papers}
\end{figure}

The application of machine learning to systems optimization has grown substantially in recent years~\citep{Wu_2022, ashouri2018survey, trofin2021mlgomachinelearningguided, van2021inquiry, reddi2025architecture2}. We collect more than 7,800 publications, showing that AI for Systems publications expanded more than 20-fold between 2017 and 2025, as illustrated in Figure~\ref{fig:ml_systems_growth_papers}. This rapid growth reflects both the increasing complexity of modern system design spaces and the maturation of learning-based techniques such as large language models and reinforcement learning that are capable of addressing systems-specific challenges. At the same time, progress has outpaced the community's ability to systematically evaluate advances, compare methods across domains, and identify shared limitations.

This survey takes a cross-stack perspective. Rather than focusing on a single subdomain, we examine how generative AI is being applied across the entire computing stack, from software engineering and distributed systems through hardware architecture to register-transfer level (RTL) design, physical layout, and verification. By looking across these layers, we find that the same structural challenges and design principles recur throughout the stack, but remain invisible when each layer is studied in isolation. Existing surveys, while valuable, typically focus on individual subdomains such as compiler optimization, computer architecture, or chip and EDA design~\citep{Wu_2022, HuangEtAl2021TODAES, PanEtAl2025LLM4EDA, GongEtAl2025LM4CodeOptSurvey, JiangEtAl2025TOSEMCodeGenSurvey, FangEtAl2025CircuitFoundationModels, HeEtAl2025FnTLLM4EDA}. This fragmentation limits the transfer of techniques between domains and prevents the identification of shared challenges in datasets, evaluation, and deployment. The central contribution of this survey is to bridge these silos: we distill five recurring challenges and five design principles that emerge from a systematic cross-layer analysis and that we believe can guide future research across the field.

Historically, computing systems optimization has relied on a combination of human expertise and algorithmic search~\citep{Frigo1998FFTW, whaley2001atlas, Puschel2005SPIRAL, ansel2014opentuner, Chen2018AutoTVM, Zheng2020Ansor}. Compiler developers design optimization passes based on decades of experience, hardware architects explore design alternatives using carefully tuned heuristics, and chip designers iteratively refine layouts through manual effort supported by incremental automation. As systems have grown more complex, spanning heterogeneous processors, deep memory hierarchies, and billion-transistor designs, this paradigm has encountered hard scalability limits. Design spaces grow combinatorially, interactions across layers defy intuition, and optimization cycles increasingly stretch to months or even years~\citep{hennessy2019new, Esmaeilzadeh2011DarkSilicon, Olson2022HeteroMemGuidance, zhang2020optimal, nardi2019practicaldesignspaceexploration, Wang2025DisaggregatedMemorySurvey, krishnan2022automatic, sampson2015hardware, wan2025generative, NVIDIA2022Hopper, Bailey2025VerificationStateSpace}.

Generative AI presents an alternative to this traditional approach. Large language models can synthesize executable code from natural language descriptions~\citep{chen2021evaluatinglargelanguagemodels, liu2023codegeneratedchatgptreally, minaee2025largelanguagemodelssurvey, wang2025efficient}, while graph neural networks can predict hardware performance with far lower cost than detailed simulation~\citep{wu2022hlsgnn, guo2022timing, zhang2020grannite, fang2023masterrtl}. Reinforcement learning agents have shown promising results in chip layout optimization, in some cases approaching or exceeding the quality of experienced human designers~\citep{mirhoseini_graph_2021}, though reproducibility remains an active area of investigation (see Challenge~3 below). Together, these results suggest that generative models trained on large collections of code, hardware traces, and design data could automate or significantly accelerate optimization across the computing stack.

Despite this promise, the field remains at an early stage. Many learned models struggle to generalize beyond their training distributions when faced with new architectures, emerging workloads, or unseen design constraints~\citep{Chen2018AutoTVM, mao2019learning, krishnan2019learningoptimizejoinqueries, kraska2018case, bachfischer2022testingrobustnesslearnedindex}. Ensuring correctness and robustness becomes more challenging as traditional formal verification techniques become less directly applicable~\citep{seshia2022formal, dreossi2019verifai, katz2017reluplex, wan2024towards}. In addition, the community lacks standardized datasets, benchmarks, and evaluation methodologies that would enable rigorous and reproducible comparison of approaches. Addressing these foundational issues will be critical in determining whether generative AI becomes a reliable tool for systems optimization or encounters the brittleness that has limited earlier ML-based efforts.

\begin{figure}[t!]
    \centering
    \includegraphics[width=0.9\textwidth]{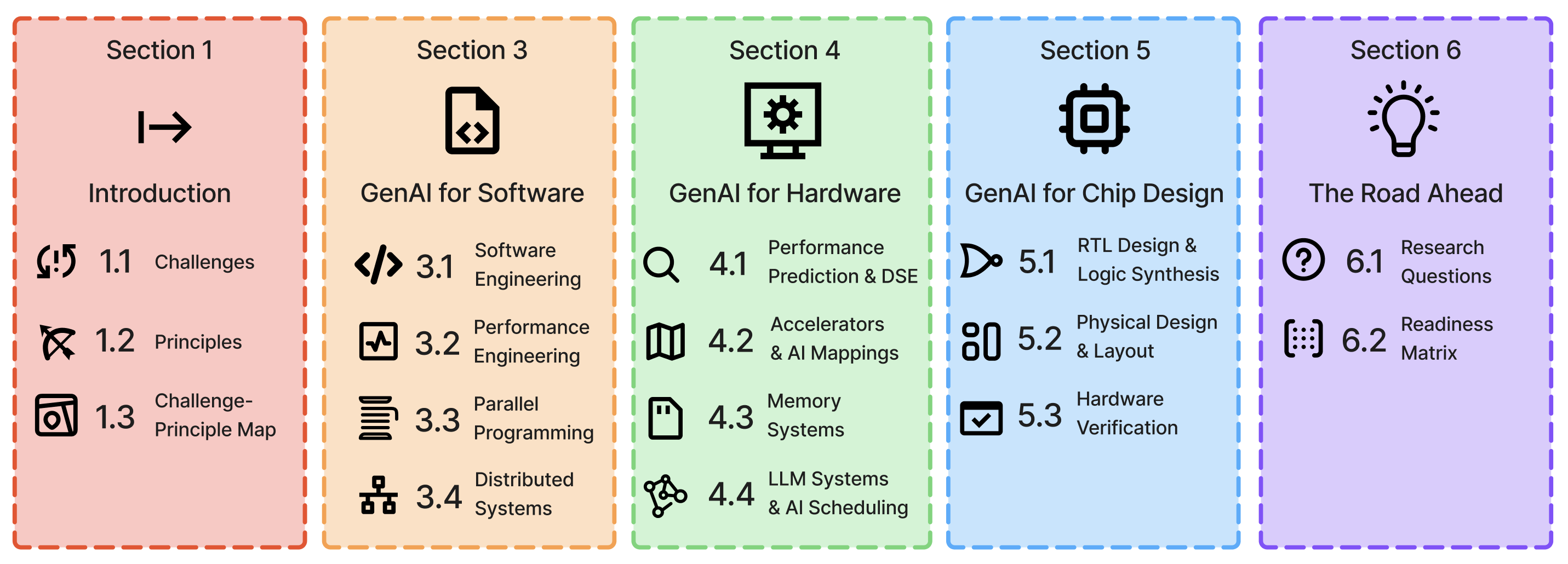}
    \caption{Scope and organization of the survey. We examine generative AI across three layers of the computing stack (software, hardware architecture, and chip design) and synthesize cross-stack challenges, design principles, and open research questions that emerge from a unified analysis.}
    \label{fig:genai_taxonomy}
\end{figure}

Concretely, we examine generative AI techniques across three major layers of the computing stack, as shown in Figure~\ref{fig:genai_taxonomy}: software (code generation, performance optimization, GPU kernels, and distributed systems), hardware architecture (performance prediction, design space exploration, accelerator design, memory systems, and workload scheduling), and chip design (RTL synthesis, physical layout, and verification)~\citep{kahng2018machine, jiang2020engineering, mehradfar2025falconmlframeworkfully, graeb2010analog, SemiEng2022ECO}. We intentionally emphasize cross-layer synthesis over exhaustive coverage within any single subdomain. Topics such as analog and mixed-signal design, neural architecture search, and the broader societal and environmental impacts of AI-driven automation fall outside our scope; we note these boundaries explicitly in Section~\ref{sec:limitation}.

For each layer, we analyze four recurring dimensions: the datasets and benchmarks used for training and evaluation, the algorithms and methods that have been proposed, the extent to which approaches have been deployed in real-world settings, and the opportunities that remain underexplored. Rather than attempting to catalog every published result, we emphasize synthesis. Compiler researchers rarely draw from advances in chip design, while hardware architects often overlook ideas from software optimization. Yet a cross-layer analysis reveals clear parallels: similar generative models applied to combinatorial search problems, recurring limitations in available datasets, and common difficulties related to verification and interpretability. The challenges and principles we distill below make these parallels explicit.

\paragraph{Who this survey is for and how to read it.}
This survey is intended for three audiences: researchers entering the field who need a structured orientation across layers, practitioners deciding which generative AI techniques to adopt for specific systems problems, and those seeking to understand where the field is heading and what open problems remain. The paper is intentionally long and cross-cutting. Readers seeking a high-level orientation may focus on the synthesis of cross-stack challenges and principles (Section~\ref{sec:challenges_principles}) and the accompanying figures, which provide a navigational overview of the design space. Readers interested in specific layers may jump directly to the corresponding sections, using the synthesis as a guide to interpret recurring patterns and tradeoffs.

\subsection*{Challenges, Principles, and Open Research Questions}
\label{sec:challenges_principles}

The survey spans a wide range of layers, from agentic software engineering and performance optimization to distributed runtimes, memory systems, LLM serving, and the full hardware toolchain from architecture exploration to RTL, physical design, and verification. This breadth is not just coverage. It reveals that many of the hardest problems are not specific to one layer, or even to one community. They are recurring structural issues that reappear whenever we place generative models inside design and decision loops. Moreover, the responses that work also converge. Despite the diversity of domains, tools, and communities, the field keeps arriving at a small number of effective approaches. 

We distill these cross-layer regularities into three views: (1) five challenges that consistently limit progress, (2) five design principles that independently emerge across layers as effective responses, and (3) open research questions that become visible only when we connect evidence across the stack. We present the challenges and principles first, then show how they connect through a challenge and principle map that can serve as a diagnostic and design aid. Table~\ref{tab:cross_stack_challenges} identifies challenges and key design principles across the computing stack.

\begin{table*}[t!]
\centering
\small
\setlength{\tabcolsep}{6pt}
\renewcommand{\arraystretch}{1.18}
\caption{Cross-stack challenges and design principles for applying Generative AI across the systems stack (Sections~\ref{sec:software}-\ref{sec:chipdesign}).
The table highlights recurring challenges (C1-C5) and the most relevant principles (P1-P5) that guide effective and trustworthy deployment.\vspace{5pt}}
\label{tab:cross_stack_challenges}

\begin{tabular}{
>{\raggedright\arraybackslash}p{0.17\textwidth}
p{0.09\textwidth}
p{0.08\textwidth}
p{0.54\textwidth}
}
\toprule

\textbf{Subsection} & \textbf{Challenges} & \textbf{Key\newline principles} & \textbf{Description} \\
\midrule

\textbf{3.1 Software engineering} &
C1, C2, C3 &
P1, P2, P3 &
Generative and agentic systems that produce code edits, patches, and refactorings in large repositories, evaluated through build systems, test suites, and CI pipelines where implicit project conventions and non-local correctness constraints dominate outcomes. \\

\textbf{3.2 Performance engineering \& code optimization} &
C1, C2, C4 &
P2, P4, P5 &
Learning-driven techniques for optimizing code, compiler flags, and program transformations using profiling feedback, where performance gains depend on hardware-specific behavior and tight coupling between compilers, runtimes, and microarchitectures. \\

\textbf{3.3 GPU kernels} &
C1, C3, C4 &
P2, P4 &
Automated synthesis and tuning of GPU kernels (e.g., CUDA, Triton), where correctness and performance are validated via expensive compile-run loops and are highly sensitive to memory access patterns, numerical behavior, and runtime interfaces. \\

\textbf{3.4 Distributed systems} &
C3, C4, C5 &
P1, P2, P3 &
Generative approaches to producing system configurations and policies for scheduling, networking, and collective communication, where changes affect live deployments and must remain safe under topology constraints and dynamically varying workloads. \\

\midrule

\textbf{4.1 Performance prediction \& DSE} &
C1, C2, C4 &
P1, P2, P4 &
Surrogate modeling and generative search methods for architectural design-space exploration, where expensive simulation or measurement limits feedback and proxy metrics encode tacit assumptions about performance and efficiency. \\

\textbf{4.2 Hardware accelerators \& AI mappings} &
C2, C4, C5 &
P2, P4, P5 &
AI-assisted accelerator design and mapping techniques that determine dataflows, scheduling, and memory placement for ML workloads, tightly integrating model structure with compiler stacks and adapting to rapidly evolving architectures and models. \\

\textbf{4.3 Memory systems \& data management} &
C2, C3, C5 &
P1, P2 &
Learning-based mechanisms for cache management, prefetching, and data placement, driven by execution traces and multi-objective goals, where adaptive policies must avoid tail-latency regressions. \\

\textbf{4.4 LLM systems \& workload scheduling} &
C3, C4, C5 &
P2, P3 &
System-level scheduling and serving of large language models, including batching, routing, and KV-cache management, where tight coupling between models, runtimes, and hardware amplifies the impact of nonstationary demand and failures. \\

\midrule

\textbf{5.1 RTL design \& synthesis} &
C1, C2, C3 &
P1, P2, P5 &
Generative assistance for RTL creation, repair, and logic synthesis, where progress is gated by slow simulation and tacit design practices embedded in hardware development workflows. \\

\textbf{5.2 Physical design \& layout} &
C1, C2, C4 &
P1, P2, P5 &
Learning-based and generative techniques for floorplanning, placement, and routing, where design quality is determined by signoff metrics and strong interdependence between geometry, timing, and congestion. \\

\textbf{5.3 Verification \& advanced chip design} &
C1, C3, C4 &
P1, P3, P5 &
AI-assisted generation of tests, assertions, and specifications for hardware verification, where correctness is the final validation and outputs must integrate tightly with simulators and formal verification tools. \\

\bottomrule
\end{tabular}

\vspace{4pt}
\footnotesize
\textbf{Challenges:}
 C1 feedback loop crisis; C2 tacit knowledge problem; C3 trust and validation;
C4 co-design across boundaries; C5 from determinism to dynamism. \textbf{Principles:} P1 embrace hybrid approaches; P2 design for continuous feedback;
P3 separate concerns by role, not by tool; P4 match approach to problem structure; P5 build on decades of systems knowledge.
\end{table*}

\subsection{Challenges}

Every layer of the stack presents its own specific difficulties, from slow simulators and brittle correctness to missing datasets and siloed toolchains. Yet when viewed together, these specific difficulties reduce to five recurring challenges. Each arises independently at multiple layers, but they share a common root: the difficulty of placing generative models inside real design and decision loops.

\paragraph{Challenge 1 (C1): The Feedback Loop Crisis.}
Across the stack, the generator is increasingly fast while the evaluator remains slow, expensive, noisy, or incomplete. At the top of the stack, agentic software development depends on tight compile, test, and execution feedback, which is exactly why benchmarks have moved from isolated function tasks to repository-scale settings like SWE-bench~\citep{jimenez2023swebench} and system interaction suites like TerminalBench \citep{merrill2026terminalbenchbenchmarkingagentshard}. In performance engineering, the same crisis appears as runtime feedback latency and brittleness, motivating profiling-in-the-loop methods like PerfCodeGen and EffiLearner \citep{peng2024perfcodegenimprovingperformancellm,huang2025effilearnerenhancingefficiencygenerated}. Moving lower, the loop becomes dramatically slower. GPU kernel synthesis requires compilation, correctness checks, and performance measurement across configurations, driving verification-heavy evaluations like robust-kbench \citep{lange2025robustkbench}. In hardware design space exploration, a single labeled datapoint can require cycle-accurate simulation or physical measurement, making feedback cost the dominant limiter and motivating surrogate models like Concorde and NeuSight \citep{nasr-esfahany_concorde_2025,lee_forecasting_2025}. In RTL and electronic design automation (EDA), tool-in-the-loop methods such as AutoChip and AIvril illustrate the same pattern, where models must repeatedly call simulators to repair outputs \citep{thakur2024autochipautomatinghdlgeneration,islam2024aivrilaidrivenrtlgeneration}. The common bottleneck is not generation. It is closing the loop quickly enough that learning and refinement are feasible.

\paragraph{Challenge 2 (C2): The Tacit Knowledge Problem.}
Systems design is shaped by knowledge that is real, decisive, and hard to write down. In software engineering, repository-level tasks expose tacit norms and constraints like build systems, dependency graphs, and undocumented conventions, which are invisible in function-level datasets and drive the push toward repo-scale benchmarks \citep{jimenez2023swebench,liu2023repobenchbenchmarkingrepositorylevelcode}. In performance optimization, much of the knowledge lives in performance folklore, compiler heuristics, and microarchitectural quirks. This is why learned compiler systems like MLGO and agentic compiler tuning like Compiler-R1 are valuable, but also why their failures are difficult to debug when the model internalizes heuristics without exposing them \citep{trofin2021mlgomachinelearningguided,pan2025compilerr1agenticcompilerautotuning}. In chip physical design, where a logical circuit must be translated into a geometric layout on silicon, the persistence of half-perimeter wirelength (HPWL) as a placement proxy is itself evidence of tacit knowledge. HPWL is widely used because it is fast and correlates with downstream objectives, even though it omits congestion, the ability of all signals to meet their timing constraints, and power integrity \citep{cheng_replace_2019,lin_dreamplace_2021}. At the lowest layers, tacit knowledge also includes the unwritten practices of verification closure, engineering change order workflows that patch designs late in the cycle, and design reuse, which motivates retrieval-based methods for RTL that ground generation in trusted IP and documentation \citep{gao2024autovcodersystematicframeworkautomated,ping2025hdlcoretrainingfreeframeworkmitigating}. Across layers, the recurring difficulty is not just learning from data. It is extracting, representing, and updating the implicit expertise that historically lived in people, tool defaults, and hard-won institutional memory.

\paragraph{Challenge 3 (C3): Trust and Validation.}
As generative components move closer to correctness-critical and cost-critical decisions, validation becomes the gating factor for deployment. In GPU kernel generation, correctness is fragile and performance can be misleading, which is why evaluation is shifting toward robustness-first setups that test multiple configurations and integrate profilers and sanitizers \citep{lange2025robustkbench}. In RTL and hardware verification, the gap between compilability and correctness is already visible in benchmark results and motivates EDA-in-the-loop refinement and pipelines that translate natural-language specifications into formal assertions checkable by verification tools \citep{thakur2024autochipautomatinghdlgeneration,fang2024assertllmgeneratingevaluatinghardware}. The verification section makes the broader point explicit: the most reliable workflows do not ask models to be correct in isolation. Instead, they ask models to produce artifacts that are independently checked by formal verification engines, simulators, or proof tools \citep{Bailey2025VerificationStateSpace,shih2025flagformalllmassistedsva}. Trust also includes scientific trust. The debate around RL-based placement and subsequent work on reproducibility and open frameworks shows that even when results look strong, limited transparency can prevent a community from trusting and building on them \citep{mirhoseini_graph_2021,markov_reevaluating_2024,cheng_assessment_2023}. Across the stack, the pattern is that adoption follows verification infrastructure, not raw model capability.

\paragraph{Challenge 4 (C4): Co-Design Across Boundaries.}
Generative AI consistently breaks layer boundaries, but our abstractions, organizations, and toolchains are still largely layered. In LLM serving, end-to-end performance depends on simultaneous choices in model behavior and system policy, including how to batch requests, manage the key-value caches that store intermediate attention state, and route work across accelerators \citep{leviathan2023fast,agrawal2024vidur}. In distributed systems, similar boundary crossing appears when models generate network configurations or collective schedules that must respect topology, safety constraints, and real deployment dynamics \citep{Wang2024-iu,Schneider2024-qd}. In hardware, the strongest industrial examples come from vertical integration where cross-boundary optimization is feasible, such as Apollo and the Pathways runtime that jointly shape accelerator design and large-scale orchestration \citep{yazdanbakhsh_apollo_2021,barham2022pathwaysasynchronousdistributeddataflow}. Even within EDA, optimizing one stage in isolation is increasingly inadequate. Early decisions about where to place major blocks constrain how wires are routed between them, which in turn determines whether signals arrive within their timing budgets. A failure at any stage can invalidate upstream architectural assumptions \citep{schlichtmann_iccad2019_overview_2019}. The core difficulty is that cross-layer gains require cross-layer control, yet our feedback signals and contracts are usually local.

\paragraph{Challenge 5 (C5): From Determinism to Dynamism.}
A growing fraction of the stack is shifting from static artifacts and deterministic heuristics toward adaptive policies that change with workload, context, and time. Agentic software development already assumes continual iteration through tests and tools \citep{liu2025agentbenchevaluatingllmsagents,merrill2026terminalbenchbenchmarkingagentshard}. In scheduling and cluster management, learned estimators and policies respond to heterogeneity, burstiness, and nonstationarity that are visible in production traces \citep{clusterdata:Wilkes2011,clusterdata:Wilkes2020a}. In networks, RL-based congestion control policies adapt online, but must coexist with safety and stability expectations historically guaranteed by conservative protocols \citep{he2025llmconf}. In memory systems, modern goals are inherently multi-objective and context-bound, with policies that must adapt across the compute-heavy prompt processing and memory-bound token generation phases of LLM inference, and across co-located tenants with competing resource demands \citep{pope2023efficiently}. Even physical design is seeing a shift toward agents that iteratively refine decisions, with hybrid approaches where learning regulates classical optimization rather than replacing it \citep{macro-regulator,lin_dreamplace_2021}. Across layers, determinism is giving way to dynamism, and the main challenge becomes how to retain predictability, debugging, and accountability when system behavior is policy-driven and stateful.

\subsection{Principles}

Just as the challenges converge, so do the responses. Across the papers we surveyed spanning software, architecture, and chip design, the approaches that succeed in practice keep arriving at a similar set of design strategies, often independently and in communities that rarely cite one another. We distill these into five principles. They are not exhaustive, nor are they prescriptions for any single layer. They are the recurring patterns we observed in the literature we reviewed, patterns that the field appears to converge toward whenever generative models are made to work reliably inside real systems. Other analyses may surface additional or different principles; we offer these as a starting point grounded in the evidence we examined.

\paragraph{Principle 1 (P1): Embrace Hybrid Approaches.}
Across domains, the most robust systems combine learning with classical structure rather than replacing it. In software, hybrid systems that integrate static analysis or symbolic checks improve reliability for critical transformations~\citep{mukherjee2021neural}. In performance prediction, hybrid models that fuse analytical bounds with compact learned components show strong data efficiency and interpretability, as in Concorde and NeuSight \citep{nasr-esfahany_concorde_2025,lee_forecasting_2025}. In physical design, approaches that treat learning as a regulator over traditional optimization-based placement engines preserve stability while benefiting from adaptive guidance \citep{macro-regulator,cheng_replace_2019}. In hardware verification, LLMs are most effective when they sit alongside model checkers and formal engines, translating intent into checkable artifacts rather than attempting to directly guarantee correctness \citep{Bailey2025VerificationStateSpace,shih2025flagformalllmassistedsva}. The recurring lesson is that hybridization is not a compromise. It is the design pattern that makes learning compatible with hard constraints and expensive errors.

\paragraph{Principle 2 (P2): Design for Continuous Feedback.}
Successful applications make the feedback loop a first-class design object. Agentic programming frameworks are evaluated by their ability to plan, execute tools, and improve through repeated interaction with tests and environments \citep{merrill2026terminalbenchbenchmarkingagentshard,liu2025agentbenchevaluatingllmsagents}. Performance engineering systems increasingly treat profiling, microbenchmarks, and regression tests as part of the optimization loop, not as a final evaluation step \citep{peng2024perfcodegenimprovingperformancellm,lin2024eco}. In RTL and EDA, tool-in-the-loop methods operationalize the same idea, using simulator logs and waveforms to repair candidate designs \citep{thakur2024autochipautomatinghdlgeneration,ho2025verilogcoderautonomousverilogcoding}. In DSE, frameworks like ArchGym formalize repeated interaction between agents, cost models, and simulators, emphasizing that how we explore can matter as much as which algorithm we choose \citep{krishnan_archgym_2023}. The cross-stack insight is that generative systems do not succeed by generating once. They succeed by making iteration cheap, structured, and measurable.

\paragraph{Principle 3 (P3): Separate Concerns by Role, Not by Tool.}
As stacks become agentic, modularity shifts from code boundaries to responsibility boundaries. In software engineering, multi-agent proposals separate planning, coding, testing, and debugging into distinct roles with different tools and memory \citep{singh2024coderesearcher}. In RTL and hardware verification, systems like Nexus and PRO-V explicitly separate generation from judging, creating an architectural analogue of generator and verifier that mirrors human workflows \citep{sami2025nexuslightweightscalablemultiagent,zhao2024prov}. LLM serving provides a clean systems example of role separation across orchestration, system control, and engine-level acceleration, where different control loops operate at different time scales and with different safety requirements \citep{agrawal2024vidur,chitty2024llm}. This principle matters across the stack because it is the main way to localize errors, assign accountability, and scale complexity when models participate in decision loops.

\paragraph{Principle 4 (P4): Match Approach to Problem Structure.}
The survey repeatedly shows that method choice should follow structure, constraints, and observability. For kernel and compiler optimization, learned cost models and guided search perform well when the space is structured and evaluable, as demonstrated by tensor program tuning frameworks like AutoTVM, Ansor, and MetaSchedule \citep{Chen2018AutoTVM,shao2022metaschedule}. In contrast, synthesis-from-scratch and policy learning become more attractive when the space is too large for templates or when human priors are weak, which is why reinforcement learning appears in chip placement and circuit design \citep{mirhoseini_graph_2021,roy_prefixrl_2021}. For performance modeling, graph-structured predictors dominate when program structure is primary, while text-native regression language models become natural when logs and configurations are the most informative interface \citep{gao_runtime_2023,akhauri2025performance}. In verification, satisfiability solvers and model checkers benefit from GNN-guided heuristics that accelerate the search for proofs or counterexamples, while specification translation benefits from LLMs paired with symbolic checking \citep{10.1145/3661308,Bailey2025VerificationStateSpace}. The cross-layer insight is that there is no single GenAI method for systems. The right approach depends on what the system exposes as structure and feedback.

\paragraph{Principle 5 (P5): Build on Decades of Systems Knowledge.}
The strongest progress comes from reusing existing abstractions, benchmarks, and invariants, then extending them to support learning. Compiler systems that incorporate ML still depend on decades of compiler engineering and intermediate representations \citep{trofin2021mlgomachinelearningguided}. Physical design methods still anchor on established placement algorithms like RePlAce and GPU-accelerated differentiable formulations like DREAMPlace that encode physics-inspired objectives \citep{cheng_replace_2019,lin_dreamplace_2021}. Distributed systems work starts from robust datacenter protocols for congestion control and load balancing, then explores how learning can propose policies while traditional mechanisms enforce safety \citep{he2025llmconf}. Memory systems research continues to rely on classical notions of locality, working sets, and allocation, even when introducing learned allocators like LLAMA or learned prefetching \citep{denning_working_1968,maas_learning-based_2020,hashemi_learning_2018}. Across the stack, GenAI is productive when it is treated as an extension of systems practice, not a replacement for it.

\subsection{The Challenge-Principle Map}

\begin{figure}[t!]
    \centering
    \includegraphics[width=0.9\textwidth]{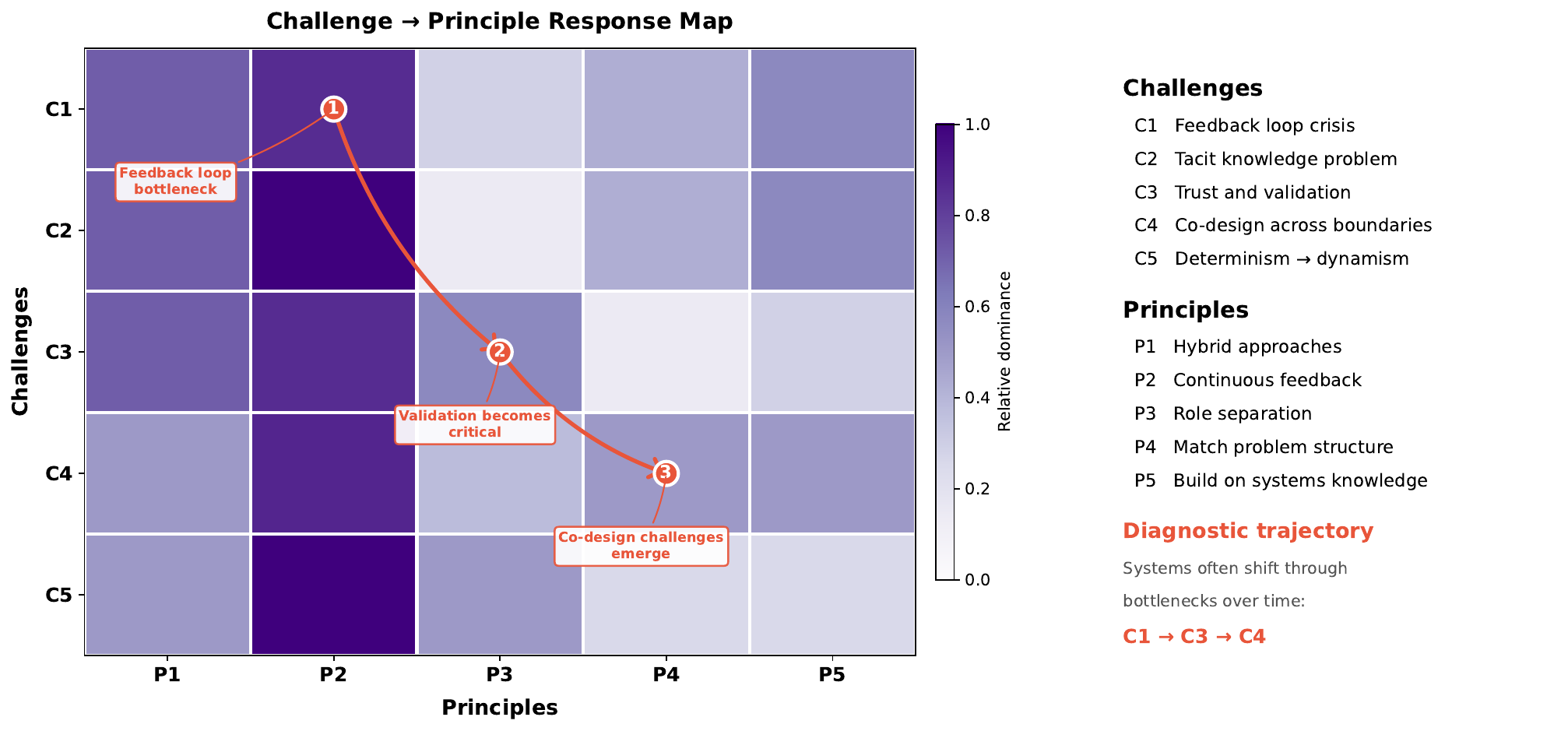}
    \caption{Challenge--principle response map with diagnostic trajectory. Rows correspond to recurring cross-stack challenges (C1--C5) and columns correspond to design principles (P1--P5). Color intensity reflects the relative dominance of each principle as a response to a given challenge, derived from the annotations in Table~\ref{tab:cross_stack_challenges}. The numbered trajectory illustrates a common maturation pattern: systems often begin with the feedback loop crisis~(C1), shift toward trust and validation~(C3) as iteration accelerates, and later expose cross-boundary co-design challenges~(C4). The map supports diagnosis and iterative navigation rather than prescribing a fixed workflow.}
    \label{fig:challenge_principle_compass}
    \label{fig:challenge_principle_navigation}
\end{figure}

With the challenges and principles now established, we can examine how they relate to one another. The fact that a wide and diverse literature reduces to five challenges and five principles is itself the central finding. It means the space of effective responses is far smaller than the space of problems, and that convergence can guide future work. Figure~\ref{fig:challenge_principle_compass} visualizes this relationship as a challenge--principle response map. Rows correspond to challenges (C1-C5) and columns to principles (P1-P5); color intensity reflects how strongly a given principle serves as a response to a given challenge, aggregated across the software, architecture, and chip design layers. The numbered trajectory overlay illustrates a common maturation pattern discussed in Example~4 below: as systems improve along one dimension, the dominant bottleneck shifts (C1~$\rightarrow$~C3~$\rightarrow$~C4), and the principles that matter shift accordingly.

This map is intended to be read as a diagnostic and design aid rather than a taxonomy. Given a system or optimization problem, one can first identify the dominant challenge it exhibits, then consult the corresponding row to see which design principles have repeatedly proven effective across layers. We illustrate this with several cross-stack examples.

\paragraph{Example 1: Feedback loop crisis (C1).}
When generation is fast but evaluation is slow or expensive, the bottleneck is not producing candidates but knowing whether they are any good. The response that works across layers is to design for continuous feedback (P2), often combined with hybrid approaches (P1) that make each iteration cheaper. In agentic software engineering, this takes the form of tight compile, test, and execution loops where the model proposes a code change, runs the test suite, observes failures, and refines its output before proposing again. In GPU kernel generation, the same logic applies: a generated kernel must be compiled, checked for correctness across input shapes, and profiled for performance, so systems like KernelBench build these steps directly into the evaluation loop rather than treating them as a final check. Lower in the stack, the cost per iteration rises dramatically. In hardware design space exploration, a single evaluation can require hours of cycle-accurate simulation, so hybrid approaches become essential. Surrogate models like Concorde and NeuSight combine analytical bounds with compact learned components to approximate expensive simulators, providing fast directional feedback that lets the search continue without waiting for full simulation at every step. The common insight is that success depends on making iteration cheap and structured rather than attempting to get the answer right in one shot.

\paragraph{Example 2: Trust and validation (C3).}
As generative components move closer to correctness-critical decisions, the question shifts from ``can the model produce a good output?'' to ``can we independently verify that the output is correct?'' The dominant response is separation of concerns by role (P3), where one component generates and a separate component judges, often combined with hybrid verification (P1) that pairs learned generation with classical checking tools. In RTL design, this plays out concretely: systems like AutoChip and VerilogCoder generate hardware descriptions, but the outputs are then fed through simulators and formal verification engines that check functional correctness independently of the model that produced them. The generator does not need to be perfect; it needs to produce artifacts that a verifier can check. The same pattern appears in GPU kernel pipelines, where generated kernels are tested not just for single-input correctness but across multiple configurations using sanitizers and profilers that catch numerical errors and performance regressions the model might miss. Trust also has a scientific dimension. The debate around RL-based chip placement showed that even strong-looking results can stall community adoption when the verification and reproducibility infrastructure is insufficient. Across the stack, the pattern is consistent: deployment follows the availability of independent validation infrastructure, not raw model capability.

\paragraph{Example 3: Co-design across boundaries (C4).}
When performance depends on decisions that span multiple abstraction layers, optimizing each layer independently leaves significant gains on the table. The response that works is to match the approach to the structure of the coupled problem (P4), designing methods that respect and exploit cross-layer dependencies rather than treating them as noise. In LLM serving, this is visible in how batching, key-value cache management, and accelerator routing interact: a scheduling decision that looks optimal from the system perspective can degrade model quality if it evicts cached state that the model needs for coherent generation, so effective systems reason about model behavior and system policy jointly. In hardware, the strongest industrial examples come from vertically integrated stacks like Google's Apollo and Pathways, where accelerator architecture and large-scale orchestration are co-designed rather than handed off sequentially. Even within chip design, the same logic applies: placement decisions constrain routing options, which determine whether timing constraints are met, so methods that treat placement and routing as independent stages consistently underperform those that account for their interaction. The core difficulty is that cross-layer gains require cross-layer feedback, yet most tools, organizations, and evaluation metrics are still structured around individual layers.

\paragraph{Example 4: Navigating the design space.}
The challenge--principle map is not static. As systems mature, improvements along one dimension often surface new bottlenecks elsewhere, and the dominant challenge shifts accordingly. The numbered trajectory in Figure~\ref{fig:challenge_principle_compass} illustrates a common progression observed across the stack, and the preceding examples make it concrete. GPU kernel generation began as a feedback loop problem (C1): early systems generated kernels but had no efficient way to evaluate them, so the field invested in compiler-and-profiler-in-the-loop pipelines that made iteration cheap. As those pipelines matured and generation quality improved, the bottleneck shifted to trust and validation (C3): a kernel that passes one test case may still fail on different input shapes or produce subtly wrong numerical results, driving the move toward robustness-first evaluation with sanitizers and multi-configuration testing. Looking ahead, the next frontier is co-design across boundaries (C4), where kernel performance depends not just on the kernel itself but on how it interacts with the surrounding memory hierarchy, scheduling policy, and workload characteristics. A similar progression is visible in chip design, where early work focused on closing the simulation feedback loop, current efforts center on building verification infrastructure that the community can trust, and the hardest remaining problems involve jointly optimizing placement, routing, and timing across traditionally siloed stages. The map captures this dynamic: solving one challenge does not end the process but reveals the next one, and the principles that matter shift accordingly.

\noindent\textbf{Using the map as a framework for systematic engineering.}
The trajectory described above is not unique to any single domain. It recurs because the underlying structure of the problem recurs: generative models are fast, evaluation is hard, trust must be earned, and real systems cross boundaries. Yet today, each community largely rediscovers these lessons independently. Kernel engineers build profiler-in-the-loop pipelines without drawing on the decades of tool-in-the-loop methodology that EDA teams have developed. Chip design teams invest in verification infrastructure without learning from the reproducibility debates that shaped software engineering benchmarks. The result is that the field advances by ad hoc construction rather than systematic engineering.

We believe the challenge--principle map points toward a more disciplined alternative. Rather than treating each new application of generative AI to systems as a greenfield effort, practitioners can use the map as a diagnostic: identify which challenge currently dominates, consult the principles that have proven effective for that challenge across other layers, and design accordingly. Concretely, this means developing shared methodologies for feedback loop design that transfer across domains, building cross-layer benchmark suites that measure not just output quality but iteration cost, validation coverage, and co-design effectiveness, and establishing best practices for when to apply hybrid approaches versus end-to-end learning based on the structure and constraints of the problem at hand. The goal is not to prescribe a fixed workflow but to give the field a common vocabulary and a systematic starting point, so that progress compounds across communities rather than being rediscovered in each one.

We return to the implications of these challenges and principles, including open research questions that become visible only from a cross-stack vantage point (Section~\ref{sec:open_questions} and Table~\ref{tab:readiness}) and the final synthesis (Section~\ref{sec:conclusion}).

\paragraph{Paper Organization.} The remainder of this survey provides the evidence base for the challenges, principles, and engineering framework outlined above. Section~\ref{sec:background} introduces background on generative AI techniques and their relevance to systems problems. Sections~\ref{sec:software}, \ref{sec:architecture}, and \ref{sec:chipdesign} examine applications at the software, architecture, and chip design layers, respectively, using the four analytical dimensions described above. In each section, readers will see the same challenges and principles reappear with layer-specific instantiations, confirming the cross-stack convergence claimed here. Section~\ref{sec:open_questions} presents open research questions that emerge from the cross-stack analysis, and Section~\ref{sec:conclusion} concludes with a synthesis of cross-cutting findings, limitations, and closing remarks.

\section{Background}
\label{sec:background}

\subsection{Computing Stack Overview}

Modern computing systems are organized as a layered stack in which high-level software expresses algorithms, compilers translate these algorithms into machine instructions, and hardware executes them through carefully designed microarchitectures implemented in silicon. At the software level, developers write and optimize programs in high-level languages that are then transformed into machine code by compilers. Software optimization spans multiple scales, ranging from low-level compiler passes that improve the performance of a single program, to tuning GPU kernels for efficient parallel execution, and up to orchestrating components in large distributed systems that support cloud services.

At the hardware architecture level, engineers design processor cores, accelerators, and memory systems that execute software efficiently. This process involves creating specialized hardware units and making architectural decisions such as cache sizes, core counts, and memory hierarchies. These choices are typically guided by performance modeling and simulation, which are used to predict how a given architecture will behave under representative workloads. Architectural design therefore serves as a bridge between software intent and physical realization, translating abstract performance goals into concrete structural decisions.

At the chip design level, the abstract architecture is implemented as a concrete circuit. Designers specify logic using register transfer level code, after which tools perform physical layout by placing and routing transistors on silicon. Extensive verification is then required to ensure that the fabricated chip behaves correctly under all expected conditions. These three layers of software, architecture, and chip design are deeply interdependent. Software decisions such as algorithm selection or memory access patterns directly affect hardware performance, while hardware capabilities such as vector units or cache hierarchies constrain the space of viable software optimizations. As a result, fully optimizing system performance and efficiency often requires a view that spans the entire computing stack.

\subsection{Traditional Machine Learning for Systems Optimization}

Over the past decade, researchers have increasingly applied machine learning techniques to assist with optimization at each layer of the computing stack. Traditional ML-for-systems approaches typically involve training models on data collected from prior design experiments or using learning agents to navigate large and complex optimization spaces. While these methods have demonstrated promising results, their effectiveness and adoption have varied significantly across different layers of the stack.

At the software layer, some of the earliest and most visible successes of machine learning in systems optimization emerged in compiler and program optimization. Learning-based techniques have been used to tune compiler heuristics and automatically select optimization passes, reducing reliance on hand-crafted rules. A survey by Ashouri~\citep{ashouri2018survey} documented many instances of compiler autotuning, including the selection of compiler flags and phase orderings that achieved substantial speedups over default configurations. More recently, Google’s MLGO framework replaced manually designed compiler decisions with reinforcement learning policies for tasks such as inlining and register allocation, yielding improved performance on real-world workloads~\citep{trofin2021mlgomachinelearningguided}.

Beyond compilers, learned models have been applied to a range of software engineering and performance optimization tasks. These include code completion, bug detection, kernel optimization, and runtime parameter tuning using techniques such as supervised learning and Bayesian optimization. Deep reinforcement learning and search-based methods have also been applied to automatically discover optimized loop transformations and tiling strategies for dense linear algebra, outperforming manually tuned libraries in specific high-performance computing domains~\citep{deng2025compilerdream}. In database systems, ML-based configuration tuning has enabled automatic adjustment of large numbers of parameters such as memory limits and indexing strategies to improve performance~\citep{van2021inquiry}. Distributed systems have also benefited, with learning-based schedulers that adaptively allocate tasks to machines under changing workloads~\citep{mao2019learning}.

Applying machine learning at the hardware architecture layer has proven more challenging. One line of work focuses on using ML models as surrogates for expensive simulations, such as predicting program runtime on a given microarchitecture. Learned models can predict basic block execution times more accurately than traditional hand-tuned analytical models. Graph neural networks have also been explored to capture hardware structure and behavior, enabling predictions of circuit timing or processor performance without full simulation.

Machine learning has also been used to navigate architectural design spaces more efficiently. Techniques such as Bayesian optimization and evolutionary strategies have been applied to explore combinations of core designs, cache configurations, and other architectural parameters more effectively than exhaustive search~\citep{nardi2019practicaldesignspaceexploration, zhang2020optimal}. Despite these advances, architecture remains a difficult domain for ML due to its inherently multi-objective nature. Architects must balance performance, power, area, and qualitative considerations such as programmability, and the optimal design often depends on workload characteristics and system-level goals. Data scarcity further complicates learning, since it is impractical to generate large numbers of real hardware designs for training. As a result, ML has achieved only niche successes in architecture, such as learned branch predictors or cache replacement policies~\citep{krishnan2019learningoptimizejoinqueries}, and progress has lagged behind that seen in software and chip design.

At the chip design and electronic design automation layer, there has been strong interest in applying machine learning to improve or automate design tasks. Physical design is a prominent example, as floorplanning, placement, and routing involve large combinatorial search problems that traditionally require weeks of expert effort. Early work used ML models to predict design quality metrics such as congestion or timing slack from partial layouts, allowing EDA tools to make better optimization decisions~\citep{kahng2018machine, jiang2020engineering}. A major milestone occurred in 2021, when a deep reinforcement learning agent was shown to perform chip floorplanning at a level comparable to or better than human experts, dramatically reducing layout time~\citep{mirhoseini_graph_2021}.

This success demonstrated that when objectives are well defined, such as minimizing wirelength or timing violations, learning-based methods can outperform manually engineered heuristics. Beyond layout, researchers have explored ML for analog circuit design, RTL power estimation, and verification tasks. In verification, learned models can help predict likely bug locations or generate effective test stimuli, augmenting traditional flows~\citep{seshia2022formal, dreossi2019verifai}. At the same time, practical deployment has revealed limitations, particularly poor generalization to novel designs or new technology nodes~\citep{bachfischer2022testingrobustnesslearnedindex}. Given the critical importance of correctness in hardware, ML methods have therefore often been used as assistive tools rather than fully autonomous solutions.

Overall, traditional machine learning approaches have delivered important but often narrow gains in systems optimization. They tend to perform best when objectives are clearly defined and sufficient training data or simulation experience is available, such as tuning a specific compiler pass or optimizing a known hardware block. These methods have been less effective for open-ended design problems that require broad generalization. As a result, the software and chip design layers saw faster adoption of ML due to abundant data and repeatable workflows, while architecture-centric tasks remained more dependent on human expertise and judgment. Many of the recurring challenges identified in Section~\ref{sec:challenges_principles}, including slow feedback loops, tacit knowledge that resists formalization, and the difficulty of validating learned outputs, were already visible in this earlier generation of work.

\subsection{Generative AI for Systems Optimization}

Generative AI refers to a class of machine learning techniques that can produce novel outputs such as text, code, images, or designs rather than simply making predictions. Recent advances in large-scale models have driven renewed interest in this paradigm, particularly through the development of foundation models that are trained on diverse and extensive datasets~\citep{brown2020gpt3}. These models learn high-level abstractions that enable them to be adapted to a wide range of downstream tasks. In the context of systems optimization, generative AI enables automation of tasks that involve synthesis or creation, such as generating optimized code, proposing hardware designs, or suggesting system configurations that have not been explicitly explored before.

Most generative AI systems follow a two-stage workflow consisting of pre-training and adaptation. During pre-training, a model is trained on a massive corpus of general data using self-supervised objectives, such as predicting the next token in large collections of text or code. This stage equips the model with broad knowledge of patterns, structures, and conventions. The model is then fine-tuned or adapted using smaller, task-specific datasets to specialize it for a particular domain or application~\citep{liu2023codegeneratedchatgptreally, minaee2025largelanguagemodelssurvey}. In systems optimization, this approach allows a model trained on general-purpose code to be adapted for tasks such as GPU kernel optimization or parallel programming. The combination of a strong prior from pre-training and targeted adaptation has enabled models to generate complex code and design artifacts that were previously achievable only through expert human effort~\citep{chen2021evaluatinglargelanguagemodels}.

Beyond one-shot generation, there is increasing interest in agentic approaches that embed generative models within closed-loop optimization processes~\citep{reddi2025architecture2}. In these settings, a model iteratively proposes solutions, evaluates feedback from simulations or tools, and refines its outputs over multiple steps. This paradigm combines ideas from reinforcement learning, planning, and generative modeling. In systems optimization, an agentic model might generate a candidate hardware design, simulate its performance, and iteratively modify the design until performance goals are met. Early examples include reinforcement learning agents that incrementally construct chip layouts through sequential placement decisions~\citep{mirhoseini_graph_2021}, as well as automated kernel synthesis frameworks that treat code generation as an iterative search process guided by hardware-in-the-loop performance feedback~\citep{ouyang2025kernelbench, ye2025flashinferefficientcustomizableattention}. By using foundation models as the core reasoning component, such agents can leverage learned knowledge while adapting dynamically based on feedback.

Generative AI techniques also extend beyond text and code to handle multiple data modalities and domain-specific representations. In software, models such as Codex translate natural language descriptions into executable programs, effectively bridging human intent and programming languages~\citep{chen2021evaluatinglargelanguagemodels}. In hardware and EDA, recent work has explored generative models that operate on circuit netlists, hardware description languages, or schematic representations~\citep{FangEtAl2025CircuitFoundationModels, HeEtAl2025FnTLLM4EDA}. Multimodal models that combine text with other signals, such as execution traces or performance metrics, enable richer contextual understanding. This capability is particularly valuable in systems optimization, where effective decisions often depend on integrating information from documentation, code, and hardware specifications.

While Transformer-based language models have received the most attention, other generative techniques are also relevant for systems. Diffusion models, which have achieved remarkable success in image generation, are being explored for tasks such as chip layout synthesis by iteratively refining designs toward high-quality solutions. Structured state space models offer an alternative approach to sequence modeling that may better handle very long sequences, such as execution traces or large netlists, where long-range dependencies are important. Together, these approaches illustrate that generative AI encompasses a diverse set of modeling tools that can be matched to the requirements of different systems problems.

Generative AI differs from earlier ML approaches in systems optimization in a key respect. Traditional techniques typically focused on improving or replacing individual heuristics within existing toolchains. In contrast, generative models aim to create complete solutions from high-level specifications, such as generating an entire parallel program or synthesizing a hardware design that satisfies given constraints. By drawing on knowledge distilled from large datasets, generative AI can address problems that were previously too complex or ill-specified for automation, opening new opportunities for optimization across the computing stack.

In the following sections, we examine how these generative AI techniques are being applied to software, hardware architecture, and chip design. For each layer, we analyze the datasets, methods, deployment experiences, and open opportunities, using the cross-stack challenges and design principles introduced in Section~\ref{sec:challenges_principles} as a recurring lens for interpretation.

\section{Generative AI for Software}
\label{sec:software}

\subsection{Software Engineering}

Agentic AI for code generation represents the top of the computing stack, where optimization shifts away from hardware constraints and toward functional specification and developer intent. Rather than focusing primarily on raw performance, modern systems emphasize correctness, maintainability, and alignment with user requirements. Autonomous agents increasingly generate, test, and refine code through iterative feedback loops, integrating execution results, tool outputs, and external context. This section reviews the datasets, algorithms, and deployment experiences that define progress in software engineering, with an emphasis on how generative AI is evaluated when the primary goal is whether software fulfills its intended purpose.

\subsubsection{Datasets and Benchmarks}

Evaluation in software engineering has evolved from isolated code synthesis toward large-scale, context-rich benchmarks that reflect real development workflows. Early datasets such as HumanEval~\citep{chen2021evaluatinglargelanguagemodels}, APPS~\citep{hendrycks2021measuringcodingchallengecompetence}, and MBPP~\citep{austin2021programsynthesislargelanguage} focused on function-level generation, where models were asked to produce small, self-contained programs that passed predefined tests. Performance on these benchmarks saturated quickly, motivating a shift toward tasks that require broader context and longer reasoning horizons. Repository-scale benchmarks such as SWE-bench~\citep{jimenez2023swebench} and RepoBench~\citep{liu2023repobenchbenchmarkingrepositorylevelcode} evaluate end-to-end changes within large codebases, requiring models to navigate file dependencies, respect existing interfaces, and satisfy build and test constraints.

\begin{figure}[t]
    \centering
    \includegraphics[width=0.85\linewidth]{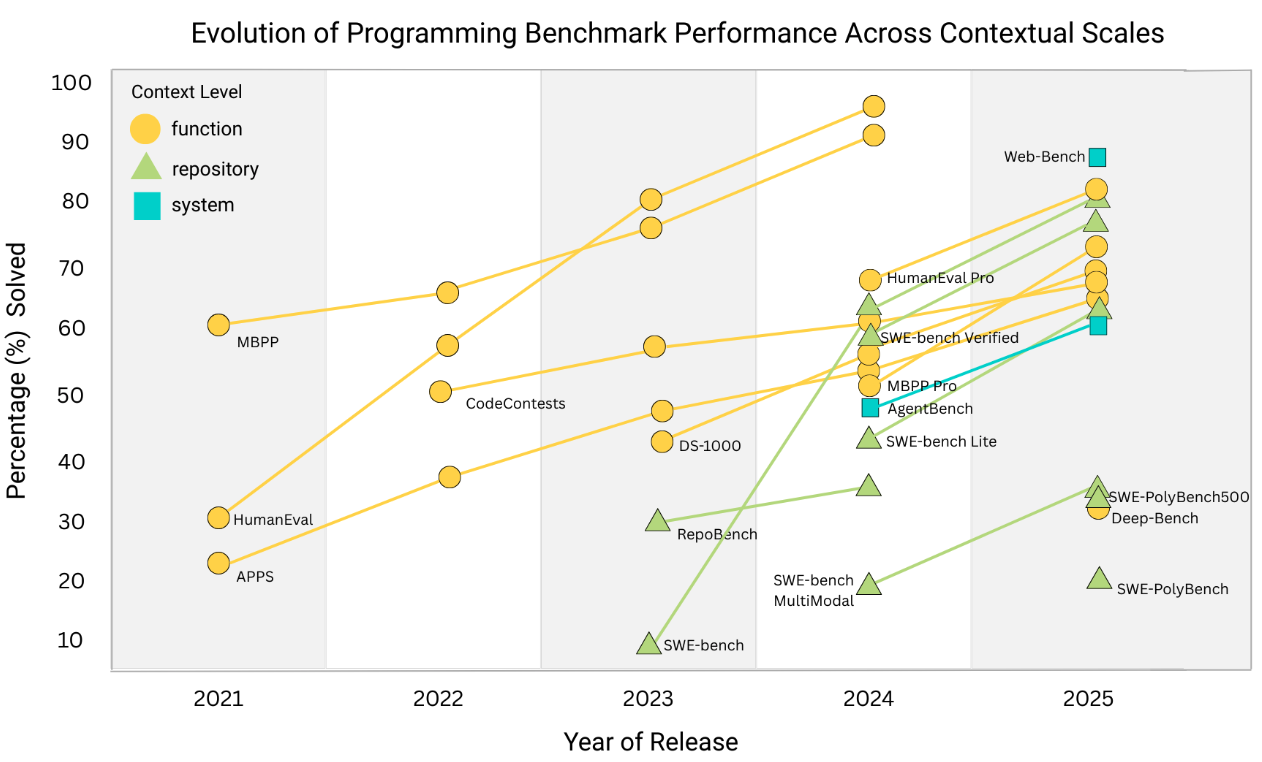}
    \caption{Benchmark progression from 2021 to 2025, illustrating the shift from function-level code generation to repository-scale and agent-based evaluation.}
    \label{fig:swe_benchmarks}
\end{figure}

More recent benchmarks emphasize agentic and system-level behavior, where models must reason across multiple steps and interact with external tools. AgentBench~\citep{liu2025agentbenchevaluatingllmsagents} and TerminalBench~\citep{merrill2026terminalbenchbenchmarkingagentshard} assess the ability to plan sequences of actions that include shell interaction, debugging, and iterative refinement. Domain-specific benchmarks such as CAD-RL~\citep{niu2025cadrl} and SIMCODE~\citep{ahmed2025simcode} introduce engineering constraints that require validation of geometric or simulation correctness. Together, these benchmarks demonstrate a pattern of rapid benchmark turnover, where tasks are quickly solved and replaced by more complex evaluations that better reflect real-world software engineering challenges, as illustrated in Figure~\ref{fig:swe_benchmarks}.

\subsubsection{Algorithms and Methods}

Algorithmic approaches for software engineering have progressed from simple text completion toward structured, closed-loop generation. Large foundation models such as Code Llama~\citep{roziere2023code} and StarCoder~\citep{li2023starcodersourceyou} serve as the backbone of many systems, leveraging extensive pretraining on diverse code repositories to capture syntax, idioms, and high-level program structure. These models provide strong priors that enable generalization across languages and problem domains, forming the basis for more advanced reasoning and planning mechanisms.

Building on these models, search and optimization techniques have been introduced to improve reliability and exploration. Methods such as MCTS-OPS~\citep{mctsops} frame code generation as a search problem, using tree-based exploration to evaluate multiple candidate solutions before selecting an output. Reinforcement learning approaches further extend this paradigm by optimizing directly for task-specific objectives beyond likelihood. Frameworks such as D-LIFT~\citep{dlift} and CAD-RL~\citep{niu2025cadrl} optimize for properties such as test success, constraint satisfaction, or geometric validity. Self-improving agentic systems such as the adaptive framework for ML library development~\citep{zhang2025adaptive} demonstrate how iterative refinement can enhance code generation quality over time. Hybrid systems combine neural generation with symbolic reasoning or static analysis to enforce correctness guarantees, which integrate learned models with program analysis tools in safety-critical settings ~\cite{mukherjee2021neural}.

\subsubsection{Real-World Deployment}

Generative AI systems for software engineering have rapidly transitioned from research prototypes to production tools. Developer-facing systems such as GitHub Copilot~\citep{fan2023largelanguagemodelssoftware} and Cursor~\citep{cursor2024} integrate directly into development environments, offering context-aware code generation, refactoring assistance, and documentation support. In these workflows, generative models function as collaborative assistants that operate continuously alongside human developers rather than as isolated code generators.

In enterprise environments, agentic systems are increasingly used to automate more complex software workflows. Systems such as AutoRAN~\citep{maxenti2025autoran} apply generative models to infrastructure configuration and management tasks, while tools like Code Researcher~\citep{singh2024coderesearcher} autonomously propose patches, execute tests, and validate changes across large repositories. These deployments demonstrate the growing trust placed in agentic coding systems while also highlighting challenges related to reliability, accountability, and integration with existing software engineering practices.

\subsubsection{Opportunities}

Several research opportunities remain open for applying generative AI to software engineering at scale. Coordinated multi-agent systems offer a promising direction for decomposing complex tasks such as planning, implementation, testing, and debugging across specialized agents that share context and intermediate state. Such architectures could improve robustness and efficiency by allowing agents to focus on distinct aspects of the development lifecycle.

Deeper integration with lower layers of the computing stack also presents significant potential. Incorporating feedback from compilers, profilers, and runtime systems would enable code generation that accounts for system-level constraints rather than focusing solely on functional correctness. Finally, the rapid saturation of existing benchmarks underscores the need for dynamic evaluation frameworks that evolve alongside model capabilities, ensuring that progress in software engineering remains measurable and meaningful over time.

\begin{takeawaybox}
\textbf{Software Engineering}
\begin{itemize}
    \item \textbf{Evolution toward repository-scale evaluation.} Benchmarks have transitioned from isolated function-level synthesis tasks, such as HumanEval, to complex environments like SWE-bench and RepoBench that require models to navigate file dependencies, respect existing interfaces, and satisfy build constraints within large codebases.
    \item \textbf{Closed-loop search and optimization algorithms.} Moving beyond simple text completion, recent approaches frame code generation as a search problem using methods like MCTS-OPS and D-LIFT to iteratively refine solutions, optimize for specific constraints, and integrate feedback from execution or static analysis tools.
    \item \textbf{Autonomous agentic workflows and integration.} Production systems are shifting from collaborative assistants like GitHub Copilot to autonomous agents capable of complex infrastructure management and patching, such as AutoRAN, with future research targeting deeper integration with compilers and profilers to address system-level constraints.
    \item \textbf{Cross-stack patterns.} This section most directly illustrates the feedback loop crisis (C1) and the tacit knowledge problem (C2), with effective responses grounded in continuous feedback (P2) and building on existing systems knowledge (P5). See Section~\ref{sec:challenges_principles}.
\end{itemize}
\end{takeawaybox}

\subsection{Performance Engineering and Code Optimization}

Performance engineering and code optimization focus on improving runtime efficiency, resource utilization, and scalability while preserving program correctness. Unlike general software engineering, optimization tasks must balance functional behavior with low-level concerns such as memory access patterns, instruction scheduling, and hardware-specific constraints. Recent advances in generative AI have enabled models to propose performance-improving code transformations, tune compilation pipelines, and synthesize optimized kernels. This section examines how progress in this area is evaluated, the methods used to achieve performance gains, and the extent to which these techniques have been deployed in real systems.

\subsubsection{Datasets and Benchmarks}

Benchmarks for LLM-driven performance engineering evaluate model capabilities across a diverse set of optimization tasks that vary in scale and realism. Some datasets emphasize algorithmic efficiency in constrained environments, such as competitive programming benchmarks like PIE~\citep{shypula2023learning}. These benchmarks test whether models can generate asymptotically efficient solutions, but they often abstract away the complexity of real-world software systems. Other benchmarks focus on isolated functions, including Mercury~\citep{du2024mercury}, EffiBench~\citep{huang2024effibench}, and EvalPerf~\citep{evalperf}, which evaluate runtime performance improvements on standard algorithms under controlled conditions.

\begin{figure}[!t]
    \centering
    \includegraphics[width=\linewidth]{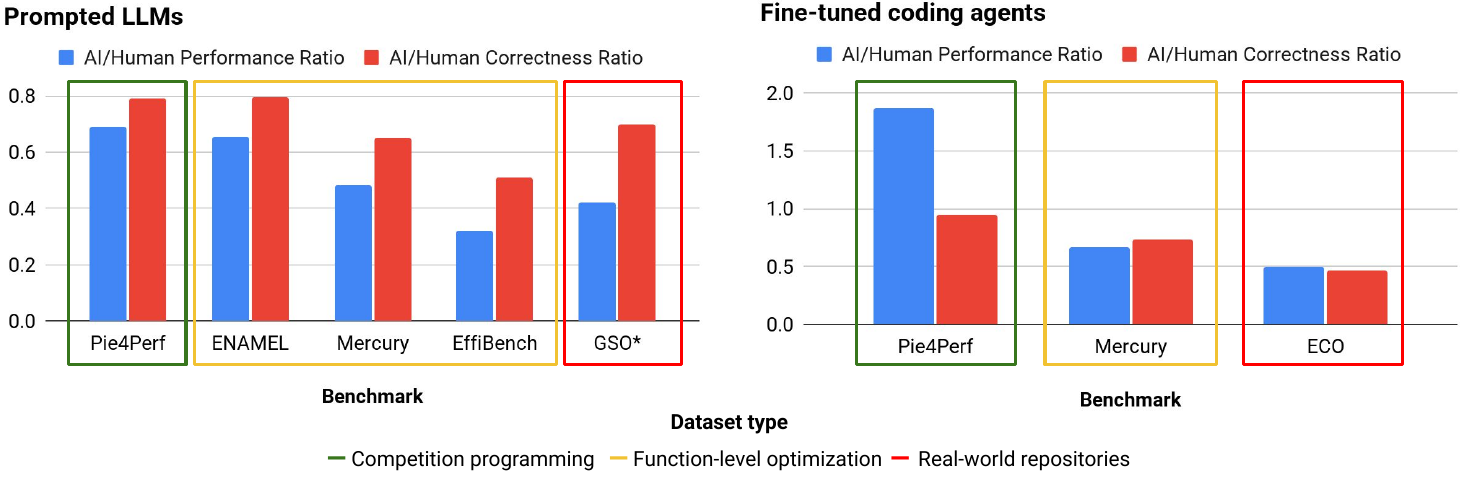}
    \caption{Large language models approach expert-level performance on isolated optimization tasks but lag on real-world repository-level optimizations. The figure compares prompt-based and fine-tuned models across benchmark categories.}
    \label{fig:Performance Engineering}
\end{figure}

To better capture production challenges, repository-level benchmarks have emerged that reflect the complexity of large codebases. Datasets such as SWE-fficiency~\citep{swefficiency}, GSO~\citep{shetty2025gso}, and SWE-Perf~\citep{swe-perf} mine performance-related code edits from open-source repositories, while ECO~\citep{lin2024eco} evaluates optimization within Google’s internal codebase. These benchmarks require models to reason over dependencies, maintain correctness across modules, and respect build and testing constraints. As shown in Figure~\ref{fig:Performance Engineering}, while current models perform well on isolated or algorithmic tasks, they struggle with the broader context required for large-scale system optimization, often necessitating human review to validate correctness.

\subsubsection{Algorithms and Methods}

Methods for automated performance optimization span a range of techniques that operate at different levels of the software stack. Many approaches rely on feedback-driven refinement, where models iteratively modify code based on profiling or runtime measurements. Systems such as PerfCodeGen~\citep{peng2024perfcodegenimprovingperformancellm} and EffiLearner~\citep{huang2025effilearnerenhancingefficiencygenerated} use execution feedback to guide successive optimization steps, enabling gradual performance improvements without explicit gradient updates.

Search-based and evolutionary strategies provide an alternative to direct learning-based refinement. Techniques that explore optimization spaces through structured search~\citep{gao2024searchbasedllmscodeoptimization} or evolutionary mechanisms, such as AlphaEvolve~\citep{novikov2025alphaevolvecodingagentscientific}, treat optimization as a black-box problem and evaluate candidate transformations based on measured performance. Compiler-level optimization has also benefited from learning-based methods, with systems such as MLGO~\citep{trofin2021mlgomachinelearningguided} and Compiler-R1~\citep{pan2025compilerr1agenticcompilerautotuning} using reinforcement learning to tune inlining decisions and other compilation heuristics. In addition, hardware-specific mapping techniques learn cost models to select optimal tensor implementations~\citep{Chen2018AutoTVM} or translate code efficiently across architectures~\citep{hong2024llmaidedcompilationtensoraccelerators}, enabling specialization for diverse hardware targets.

\subsubsection{Real-World Deployment}

In practice, performance optimization systems have evolved from passive analysis tools to active agents that modify code automatically. Within Google’s production environment, optimization efforts progressed from Google-Wide Profiling~\citep{gwp} to AutoFDO~\citep{chen2016autofdo}, which automated feedback-directed optimization, and more recently to ECO~\citep{lin2024eco}, which applies LLM-generated performance patches directly to production code. These systems demonstrate how generative models can be integrated into large-scale development workflows when supported by extensive testing and validation infrastructure.

Outside hyperscale environments, deployment patterns differ due to resource constraints and heterogeneity. Tools such as Galaxy~\citep{ye2024galaxy} and LLaMoCo~\citep{ma2026llamoco} apply AI-driven optimization techniques to consumer and edge workloads, focusing on heterogeneous execution across CPUs, GPUs, and accelerators. These deployments emphasize portability and usability, highlighting a contrast between centralized, data-rich optimization pipelines and distributed, user-facing optimization tools.

\subsubsection{Opportunities}

Despite significant progress, several challenges remain open in performance engineering with generative AI. A primary limitation is maintaining correctness in complex contexts, where optimizing one component can introduce subtle errors due to hidden dependencies in large repositories. Improving robustness in these settings will require tighter integration between optimization models and program analysis or verification tools.

Another key opportunity lies in addressing hardware heterogeneity more explicitly. Future benchmarks and models must move beyond generic efficiency metrics and instead generate optimizations tailored to specific hardware targets, memory hierarchies, and execution environments. Finally, the rapid improvement of models on existing benchmarks underscores the need for dynamic evaluation frameworks that evolve over time, preventing overfitting and ensuring that measured progress reflects real-world performance gains rather than benchmark-specific tuning.

\begin{takeawaybox}
\textbf{Performance Engineering and Code Optimization}
\begin{itemize}
    \item \textbf{Gap between isolated and repository-level benchmarks.} While models demonstrate expert-level proficiency on isolated algorithmic tasks like Mercury~\citep{du2024mercury}, they struggle with repository-level benchmarks such as SWE-fficiency~\citep{swefficiency} and ECO~\citep{lin2024eco} that require reasoning over complex dependencies and build constraints.
    \item \textbf{Feedback-driven refinement and compiler autotuning.} Advanced optimization methods utilize iterative execution feedback to guide code modifications in systems like PerfCodeGen~\citep{peng2024perfcodegenimprovingperformancellm}, while reinforcement learning agents in Compiler-R1~\citep{pan2025compilerr1agenticcompilerautotuning} automate low-level compilation heuristics and inlining decisions.
    \item \textbf{Production deployment and heterogeneity challenges.} Real-world implementations have evolved from profiling tools to active patching agents like Google's ECO~\citep{lin2024eco}, though significant challenges remain in ensuring correctness across large codebases and generating optimizations tailored to heterogeneous hardware targets.
    \item \textbf{Cross-stack patterns.} This section most directly illustrates the feedback loop crisis (C1) and the tacit knowledge problem (C2), with effective responses grounded in hybrid approaches (P1) and continuous feedback (P2). See Section~\ref{sec:challenges_principles}.
\end{itemize}
\end{takeawaybox}

\subsection{Parallel Programming and GPU Kernels}

Optimizing parallel programming kernels such as GEMM and attention operations is central to modern machine learning performance. Kernel efficiency directly determines end-to-end training throughput, inference latency, and energy consumption, especially as models scale in size and complexity. As a result, parallel programming kernel optimization on GPUs has long been a focal point for both hardware vendors and systems researchers. This subsection examines how datasets, algorithms, and deployment practices have evolved to support increasingly automated and hardware-aware kernel optimization.

Rather than treating kernel optimization as a static compiler problem, recent work frames it as an adaptive, data-driven process. Learned and generative methods increasingly operate in closed loops with compilers, runtimes, and hardware, enabling rapid specialization for new workloads and architectures. We organize this discussion around the benchmarks that shape evaluation, the algorithmic methods that define progress, and the extent to which these techniques have transitioned into real systems.

\begin{figure}[t]
    \centering
    \includegraphics[width=0.8\textwidth]{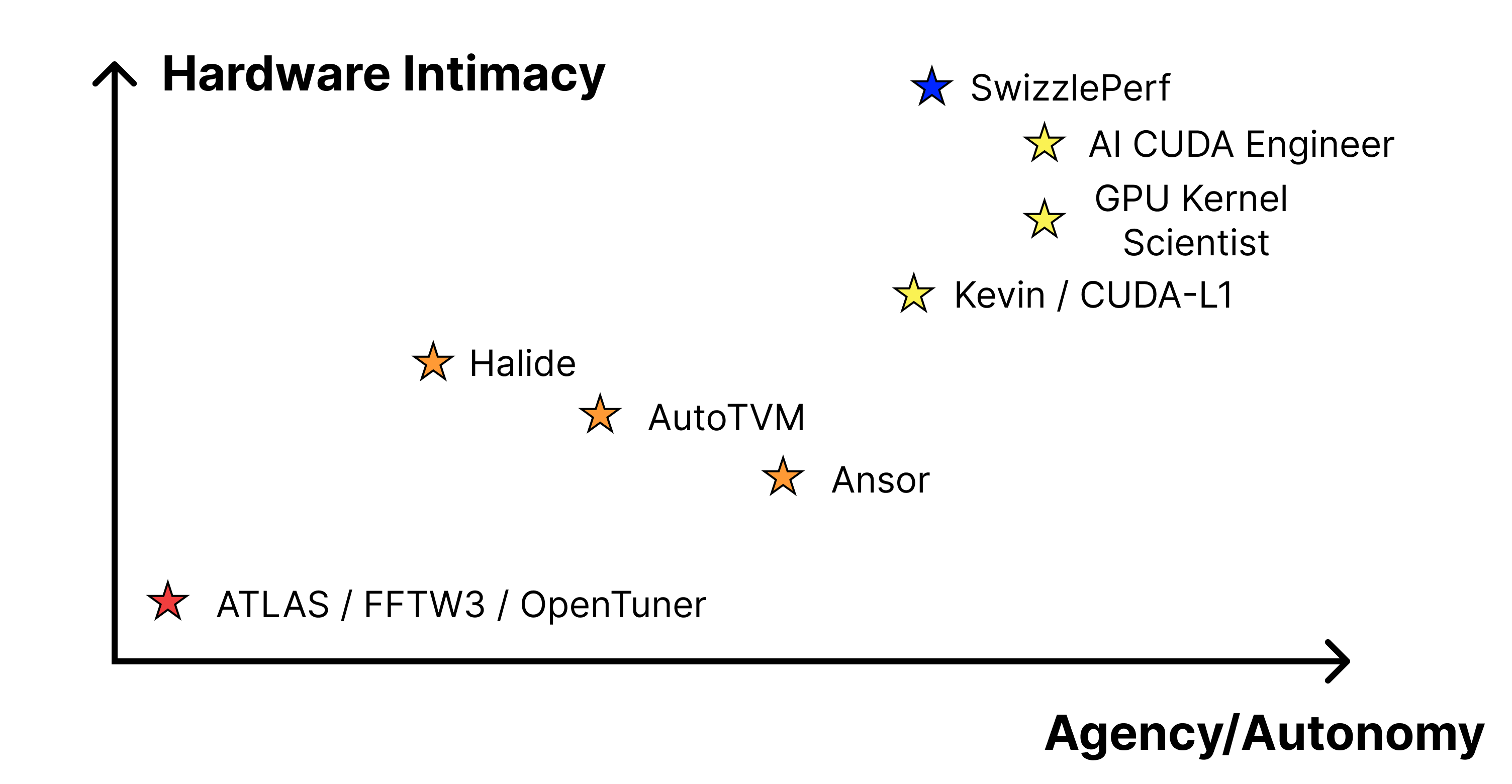} 
    \caption{Taxonomy of AI-driven GPU kernel optimization, illustrating the evolution from empirical autotuning to learned cost models and generative synthesis.}
    \label{fig:taxonomy}
\end{figure}

\subsubsection{Datasets and Benchmarks}

Benchmarks for GPU kernel optimization must simultaneously evaluate functional correctness, performance, and robustness across diverse hardware configurations. Early autotuning systems relied on implicit evaluation through compiler feedback or runtime profiling, but recent work has introduced explicit benchmarks that expose failure modes and generalization limits.

KernelBench evaluates whether language models can generate correct and performant kernels across 250 PyTorch workloads, measuring both execution correctness and speed~\citep{ouyang2025kernelbench}. While effective for assessing functional generation, single-configuration evaluation can mask numerical instability or architecture-specific failures. Robust-kbench addresses this gap by testing kernels across multiple configurations and verifying behavior using profilers and sanitizers, revealing brittleness that does not appear in narrow evaluations~\citep{lange2025robustkbench}.

Other benchmarks emphasize production realism. FlashInfer-Bench evaluates kernel behavior using traces from deployed inference systems, incorporating irregular batch sizes and dynamic shapes that are common in real workloads~\citep{ye2025flashinferefficientcustomizableattention}. To support learning-based approaches, large-scale datasets such as the AI-CUDA Archive provide approximately 30,000 kernels with rich profiling metadata~\citep{lange2024aicuda}, while KernelBook aggregates Triton and CUDA examples curated for educational and training purposes~\citep{gpumode2025kernelbook}. SwizzlePerf complements these resources with a hardware-aware evaluation suite focused on spatial swizzling patterns that stress memory subsystems~\citep{tschand2025swizzleperf}.

\subsubsection{Algorithms and Methods}

Algorithmic approaches to GPU kernel optimization have progressed through several distinct phases, corresponding to increasing levels of automation and hardware intimacy, as illustrated in Figure~\ref{fig:taxonomy}. Early systems relied on heuristic and empirical autotuning, where structured search explored parameterized templates. Systems such as ATLAS and OpenTuner generalized multi-strategy search across implementations and achieved strong performance portability, but were limited by high search cost and the rigidity of predefined templates~\citep{whaley2001atlas,ansel2014opentuner}.

To address scalability, learned cost models and scheduling frameworks were introduced. AutoTVM and Ansor trained predictive models to estimate kernel performance without executing every candidate, enabling exploration of much larger scheduling spaces~\citep{shao2022metaschedule}. MetaSchedule later unified these ideas under a probabilistic programming abstraction, providing a common framework for learning-guided search across hardware targets~\citep{shao2022metaschedule}. These approaches marked a shift from brute-force search to model-guided optimization, significantly reducing tuning overhead.

The most recent frontier moves beyond schedule selection to model-in-the-loop synthesis, where kernels are generated or transformed directly. Frameworks such as Triton~\citep{tillet2019triton} and PyTorch FlexAttention~\citep{he2024flexattention} expose programmable kernel abstractions that enable fusion and specialization at a higher level. Reinforcement learning agents Kevin~\citep{baronio2025kevin} and generative models operate within these systems to optimize low-level representations and translate high-level PyTorch code into efficient kernels. Agentic systems including Astra~\citep{wei2025astra}, AI CUDA Engineer~\citep{lange2024aicuda}, CUDA-L1 ~\citep{li2025cuda}, KernelEvolve~\citep{liao2026kernelevolvescalingagentickernel}, and AccelOpt~\citep{zhang2025accelopt} demonstrate rapid iteration and strong performance gains, but also expose new challenges related to correctness, numerical stability, and verification at low abstraction levels.

\subsubsection{Real-World Deployment}

AI-driven GPU kernel optimization is increasingly deployed in widely used machine learning frameworks. Triton provides a programmable interface that allows developers and automated systems to generate and tune custom kernels, while PyTorch FlexAttention integrates learned optimization directly into attention operators used in large models. These systems demonstrate that learned and generative approaches can coexist with traditional compiler infrastructures when embedded behind stable abstractions and validation pipelines.

In production environments, deployment remains conservative. Kernel synthesis systems are typically constrained to well-scoped components or used in advisory roles, with extensive testing and fallback mechanisms to mitigate the high cost of kernel failures. Human oversight and staged rollout remain common, particularly when kernels target diverse hardware configurations or low-level instruction sets.

\subsubsection{Opportunities}

Several research opportunities remain open in AI-driven GPU kernel optimization. Robustness-first evaluation is needed to expose irregular memory access patterns, numerical instability, and rare edge cases that are underrepresented in existing benchmarks. Learned synthesis approaches would benefit from tighter integration with verification techniques, such as symbolic contracts or intermediate representation constraints, to ensure correctness throughout the generation process. Recent work begins to address this gap~\citep{chatterjee2025proofwright}, but scalable verification remains an open challenge.

Reward design also requires further attention. Optimization objectives must account not only for latency but also for energy efficiency, determinism, and hardware utilization. Finally, connecting micro-kernel improvements to application-level metrics such as throughput and time-to-train remains essential. Without end-to-end impact tracking, kernel-level gains risk becoming isolated optimizations that fail to translate into meaningful system-level benefits.

\begin{takeawaybox}
\textbf{Parallel Programming and GPU Kernels}
\begin{itemize}
    \item \textbf{Evolution to Generative Kernel Synthesis.} Optimization strategies have progressed from empirical autotuning and learned cost models like MetaSchedule to agentic systems such as Kevin and Astra that directly generate or transform kernel code. While these generative approaches enable rapid specialization for high-level abstractions like Triton, they introduce significant challenges regarding low-level verification and numerical stability compared to traditional compiler heuristics.
    \item \textbf{Shift Toward Robustness-Aware Benchmarking.} Recent benchmarks like Robust-kbench and FlashInfer-Bench move beyond simple latency metrics to evaluate functional correctness, numerical stability, and behavior under production traces with irregular shapes. This shift addresses the limitations of single-configuration evaluations by exposing brittleness and architecture-specific failures that generative models often overlook.
    \item \textbf{Deployment Constraints and Verification Gaps.} Although frameworks like PyTorch FlexAttention and Triton have integrated learned optimization, production deployment remains conservative due to the high cost of kernel failures. Future research focuses on integrating formal verification techniques, such as those in ProofWright, and aligning micro-kernel objectives with end-to-end application metrics like training throughput and energy efficiency.
    \item \textbf{Cross-stack patterns.} This section most directly illustrates the feedback loop crisis (C1) and trust and validation (C3), with effective responses grounded in hybrid approaches (P1) and continuous feedback (P2). See Section~\ref{sec:challenges_principles}.
\end{itemize}
\end{takeawaybox}

\subsection{Distributed Systems Integration}

\begin{figure}[h]
    \centering
    \includegraphics[width=0.6\linewidth]{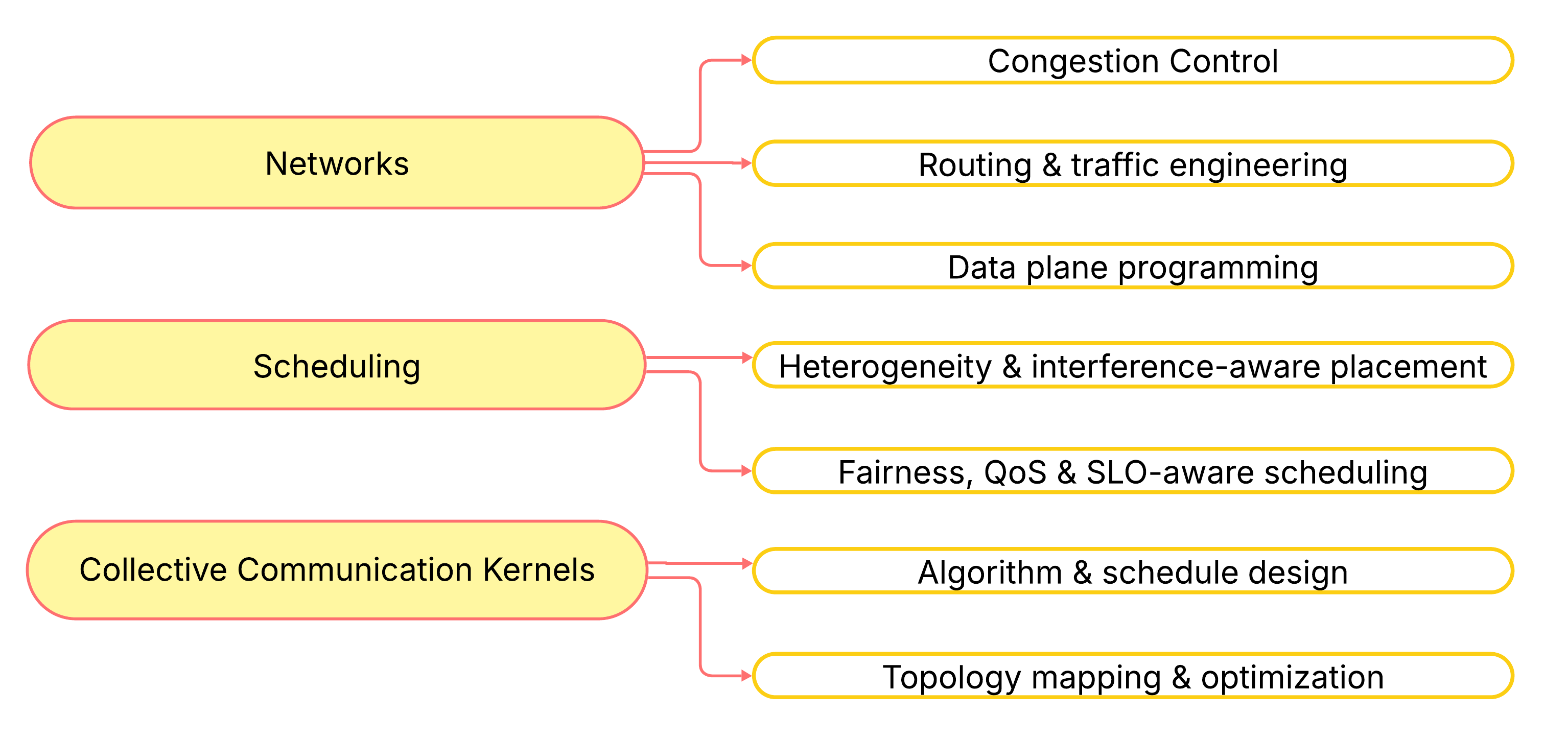}
    \caption{Taxonomy of generative AI applications in distributed system integration, spanning network control, scheduling, and collective communication.}
    \label{fig:distributed_taxonomy}
\end{figure}

Distributed systems integration addresses how large-scale computing systems coordinate computation and data across many machines while balancing performance, reliability, and hardware constraints~\citep{clusterdata:Verma2015,schwarzkopf2013omega} as shown in Figure ~\ref{fig:distributed_taxonomy}. Core challenges span network control, cluster scheduling, and collective communication, each of which involves complex policies traditionally designed through first-principles reasoning and extensive operational tuning~\citep{alizadeh2010data,ghodsi2011dominant}. Generative AI is emerging as a complementary tool that can assist with policy synthesis, configuration, and optimization across these domains~\citep{mao2019learning,Schneider2024-qd}. This section organizes recent work through a common framework of datasets, algorithms, deployment experience, and open opportunities.

\subsubsection{Datasets and Benchmarks}

Evaluation resources for generative AI in distributed systems remain limited and fragmented across subdomains. In networking, benchmarks such as NetConfEval evaluate intent-to-configuration tasks, measuring whether models can translate high-level goals into correct routing or ECMP configurations~\citep{Wang2024-iu}. Studies on P4 generation and compilation similarly highlight the need for domain-specific datasets to evaluate correctness and semantic fidelity in data plane programming~\citep{Dumitru2024-rz}. These benchmarks emphasize that smaller, domain-tuned models can outperform general-purpose LLMs when evaluated on specialized networking tasks.

For scheduling, datasets typically derive from production cluster traces or synthetic workload generators~\citep{clusterdata:Wilkes2011,clusterdata:Wilkes2020a}. Recent work introduces supervised corpora for classical scheduling problems such as job-shop scheduling, as well as large-scale log analyses from production systems that enable LLM-based copilots to surface optimization opportunities~\citep{akhauri2025performance}. In collective communication, no benchmark yet directly targets LLM-generated schedules. Related efforts such as KernelBench and ComputeEval provide partial templates~\citep{ouyang2025kernelbench,Rodriguez2025-qt}, but the lack of topology-aware, collective-specific benchmarks remains a major gap for evaluating generative approaches at scale~\citep{Schneider2024-qd}.

\subsubsection{Algorithms and Methods}

Generative AI methods in distributed systems primarily operate at the level of policy synthesis, parameter tuning, and decision support rather than direct low-level control. In networking, language models have been used to synthesize congestion control algorithms in simulation and to translate intent into device configurations or programmable data plane code~\citep{Dumitru2024-rz}. Reinforcement learning approaches, such as PCC-style controllers, demonstrate that learned policies can adapt congestion behavior to runtime conditions, although deployment typically requires strict safety constraints~\citep{Dong2018-jm}.

Scheduling research spans reinforcement learning, hybrid optimization, and LLM-assisted planning. End-to-end RL systems such as Decima learn scheduling policies for workflow graphs~\citep{mao2019learning}, while other approaches combine learned performance estimators with classical optimizers to handle heterogeneity and interference-aware placement~\citep{Delimitrou2013-paragon}. Language models increasingly act as copilots that analyze logs, propose configuration changes, or synthesize placement rules from natural-language policies~\citep{akhauri2025performance}. For collective communication, generative approaches focus on synthesizing or refining communication schedules and kernels. Systems such as MPIrigen fine-tune LLMs on communication patterns~\citep{Schneider2024-qd}, while hardware-in-the-loop reinforcement learning optimizes kernel implementations for specific topologies~\citep{Chen2025-px}.

\subsubsection{Real-World Deployment}

Deployment of generative AI in distributed systems remains cautious due to the high cost of failures and the need for predictable behavior. In networking, GenAI systems are typically integrated into validation pipelines, where generated policies are checked in simulation or emulation before deployment~\citep{Wang2024-iu}. Hybrid designs are common, with generative models proposing configurations and traditional control mechanisms enforcing safety at runtime~\citep{Zhu2015-ru}.

In scheduling, production systems continue to rely on engineered heuristics such as backfilling and dominant resource fairness~\citep{ghodsi2011dominant,zaharia2010delay}, with ML components used primarily for estimation or advisory roles~\citep{Narayanan2020-gavel}. LLM-based copilots that assist operators in diagnosing performance issues or tuning parameters are more common than fully autonomous schedulers. For collective communication, deployment has focused on learned tuning and synthesis within tightly scoped environments, with production-scale systems still dominated by carefully engineered libraries such as NCCL~\citep{Hu2025-ba}.

\subsubsection{Opportunities}

Several challenges must be addressed for generative AI to play a larger role in distributed systems integration. Robust, production-realistic benchmarks are needed to evaluate models under failure scenarios, workload diversity, and multi-tenant conditions. Hardware-in-the-loop validation remains critical, particularly for networking and collective communication, where simulation often fails to capture timing-sensitive behavior.

A promising direction is hybrid system design, where generative models handle high-level policy synthesis or intent translation, while fast, well-understood mechanisms enforce safety and correctness. In scheduling, combining learned estimators with classical optimizers appears especially deployable in the near term. Finally, scaling generative approaches to extreme environments, such as clusters with tens or hundreds of thousands of GPUs, will require advances in efficiency, topology-aware modeling, and principled interfaces between GenAI controllers and existing distributed systems infrastructure~.

\begin{takeawaybox} \textbf{Distributed Systems Integration} \begin{itemize} \item \textbf{Fragmentation and the need for topology-awareness.} While networking has established benchmarks for intent-to-configuration (e.g., NetConfEval ~\citep{Wang2024-iu}), collective communication lacks standardized, topology-aware benchmarks. Evidence suggests that specialized, domain-tuned models are more effective than general-purpose LLMs for infrastructure-specific syntax like P4 or MPI~\citep{Wang2024-iu,Dumitru2024-rz}. \item \textbf{Transition from RL-policies to Generative Planning.} The field is moving from discriminative RL-based scheduling (e.g., Decima ~\citep{mao2019learning}) toward generative agents that perform log-based diagnostic analysis and high-level policy synthesis, bridging the gap between natural language intent and operational parameters~\citep{akhauri2025performance}. \item \textbf{Hybrid Safety and Role Separation.} To mitigate the high cost of distributed failures, deployment relies on "sandboxed" architectures: generative models propose configurations or communication schedules (P3: Role Separation), which are then strictly enforced by proven mechanisms like NCCL, DRF, or formal simulation before execution~\citep{Hu2025-ba,Zhu2015-ru}. \item \textbf{Cross-stack patterns.} This section illustrates the difficulty of Co-design across boundaries (C4) and the move From determinism to dynamism (C5). The most successful responses leverage Hybrid approaches (P1) and Separation of concerns by role (P3). See Section~\ref{sec:challenges_principles}. \end{itemize} \end{takeawaybox}

\section{Generative AI for Hardware Architecture}
\label{sec:architecture}

\subsection{Performance Prediction and Design Space Exploration}

The design of modern hardware architectures represents a search and optimization problem of astronomical scale. Engineers must navigate a combinatorial space of design parameters, spanning core counts, cache hierarchies, memory controller settings, and on-chip interconnects, to identify Pareto-optimal designs that balance performance, power, and area \citep{hennessy2019new}. Traditionally, this design space exploration process has relied on slow, cycle-accurate simulators. The prohibitive cost of simulation, which can require days or weeks per design point, creates a severe data bottleneck that restricts exploration to only a tiny fraction of the full space \citep{nasr-esfahany_concorde_2025}.

This section examines how generative AI and modern learning methods address this dual challenge. We first consider performance prediction, which focuses on building fast surrogate models that estimate metrics such as cycles per instruction, latency, and throughput orders of magnitude faster than detailed simulation. We then examine design space exploration methods that leverage these predictors to search efficiently for high-quality architectures. The evolution of this area reflects a sustained response to data scarcity in hardware design, progressing from classical statistical models to neural networks and, more recently, to hybrid and generative approaches that integrate analytical priors and enable increasingly automated co-design.

\subsubsection{Datasets and Benchmarks}

Progress in performance prediction and design space exploration is fundamentally constrained by the availability of high-quality and representative datasets. Unlike domains such as computer vision or natural language processing, which benefited from large public corpora such as ImageNet \citep{Deng2009ImageNet} or web-scale text \citep{brown2020gpt3}, the hardware domain is inherently data starved \citep{Wu_2022}. Each ground-truth data point typically requires complex simulation or measurement on physical hardware, making data collection expensive and slow. As a result, the field has evolved a fragmented ecosystem of specialized benchmarks that target different layers of the architecture stack and different performance metrics.

This fragmentation is unavoidable because performance is highly context dependent. Metrics vary by domain, ranging from CPI for CPUs, to latency and throughput for accelerators, to combined power and area objectives for high-level synthesis. TpuGraphs provides large-scale graph-based representations of tensor programs with performance metrics, enabling evaluation of graph neural network predictors for accelerator workloads \citep{phothilimthana2023tpugraphs}. ForgeHLS offers more than 400,000 HLS designs annotated with performance, power, and area outcomes, emphasizing Pareto frontier exploration through Bayesian DSE \citep{peng_forgehls_2025}. PerfCastDB contributes large-scale measurements from real Intel Xeon processors paired with SPEC benchmarks, grounding prediction models in physical hardware behavior rather than simulation \citep{liu_ncpp_2024}. This diversity of datasets directly shapes the specialized algorithms developed to operate within each subdomain.

\subsubsection{Algorithms and Methods}

\begin{figure}[t]
    \centering
    \includegraphics[width=0.75\linewidth]{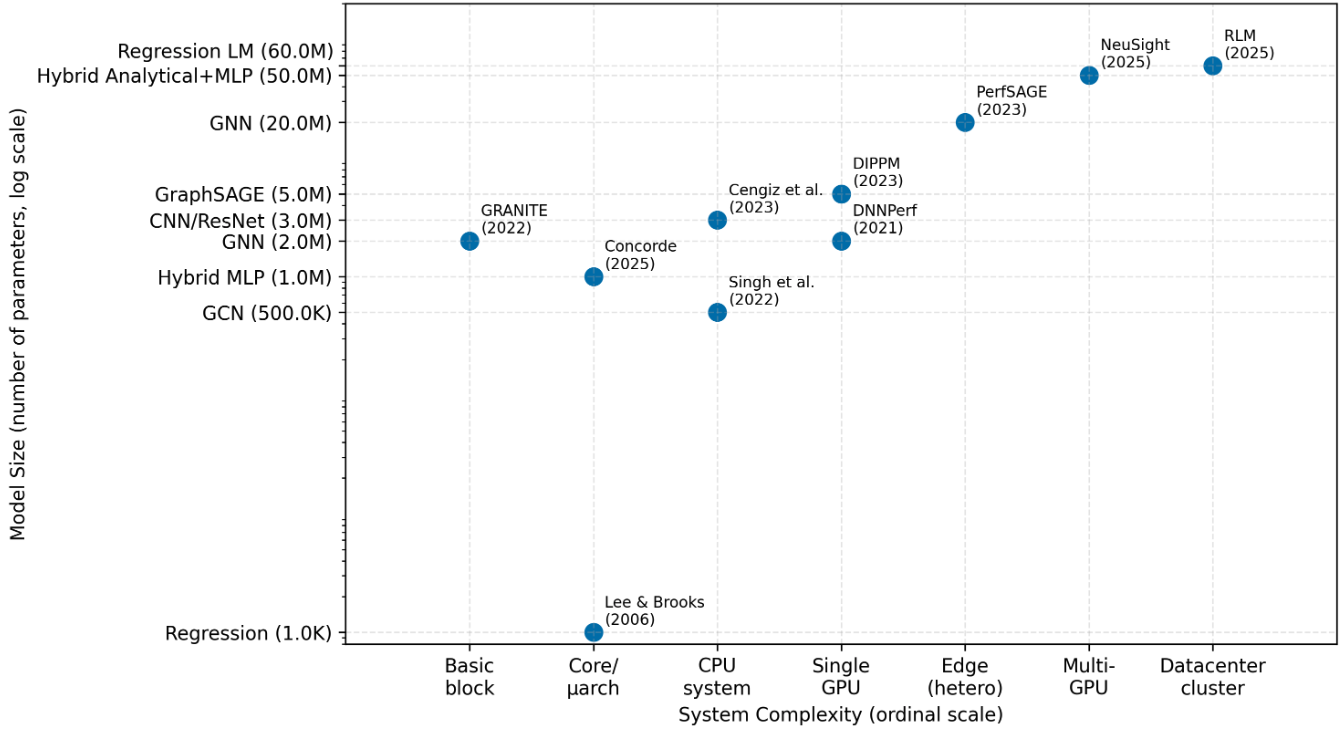}
    \caption{Relationship between computing system complexity and prediction model complexity across representative performance prediction studies. The figure illustrates the progression from early regression models to deep graph networks, hybrid analytical–ML systems, and regression language models as hardware systems grow in scale and heterogeneity \citep{gao_runtime_2023, chai_perfsage_2023, nasr-esfahany_concorde_2025, lee_forecasting_2025, sun2023gibbon, akhauri2025performance, zhang2025recursive}.}
    \label{fig:perf_prediction}
\end{figure}

Across performance prediction and design space exploration, recent progress reflects a shift toward structure-aware models and hybrid analytical–learning approaches with increasing complexity, as shown in Figure~\ref{fig:perf_prediction}. These methods explicitly encode architectural structure, resource decomposition, or physical constraints to improve accuracy, generalization, and interpretability. Unifying frameworks such as ArchGym accelerate algorithm comparison and promote reusable surrogates, while tools such as HyperMapper formalize constrained, multi-objective exploration with human priors \citep{krishnan_archgym_2023, nardi2019practicaldesignspaceexploration}. Empirical observations such as the hyperparameter lottery effect, the need to accommodate evolving toolchains, and the importance of attribution analysis underscore that careful method design often matters as much as model choice.

For performance prediction, learned surrogates increasingly replace monolithic simulators. On CPUs, Concorde combines simple per-resource analytical bounds with a compact neural combiner, enabling constant-time inference from performance distributions while preserving interpretability \citep{nasr-esfahany_concorde_2025}. On GPUs, NeuSight predicts deep learning performance by modeling tile-level behavior and constraining predictions with compute and bandwidth laws, allowing generalization to unseen devices \citep{lee_forecasting_2025}. For processing-in-memory architectures, Gibbon integrates analytical models with machine learning to jointly optimize neural networks and hardware configurations \citep{sun2023gibbon}. Graph neural networks dominate end-to-end workload prediction, with systems such as DNNPerf, PerfSAGE, and DIPPM estimating latency, energy, and memory across diverse platforms \citep{gao_runtime_2023, chai_perfsage_2023, selvam_dippm_2023}. A recent trend explores text-native modeling, where regression language models treat configurations and logs as text, achieving high accuracy and rapid adaptation with minimal feature engineering \citep{akhauri2025performance}.

Design space exploration methods increasingly favor generative and learning-based search over handcrafted heuristics. ArchGym standardizes the interface between search agents and cost models, enabling fair comparison of reinforcement learning, Bayesian optimization, and evolutionary strategies \citep{krishnan_archgym_2023}. When design parameters decompose naturally, multi-agent reinforcement learning assigns control of subsystems to separate agents, improving scalability \citep{krishnan_multi-agent_2022}. White-box optimizers such as HyperMapper 2.0 handle unknown feasibility and integrate expert priors, while black-box systems such as AutoDSE embrace toolchain variability and reduce manual tuning through bottleneck-guided exploration \citep{nardi2019practicaldesignspaceexploration, sohrabizadeh_autodse_2022}.

\subsubsection{Real-World Deployment}

Performance prediction and design space exploration have seen some of the most impactful industrial deployments of ML-for-systems techniques, particularly within vertically integrated organizations. Google demonstrated a reinforcement learning agent for chip placement that produced manufacturable TPU floorplans in hours rather than months, directly deploying generative optimization in production silicon design \citep{mirhoseini_graph_2021}. NVIDIA has reported the use of reinforcement learning to design arithmetic circuits for production GPUs, discovering novel designs that improved area and performance \citep{roy_prefixrl_2021, song_circuitvae_2024}.

Larger co-design platforms further illustrate real-world adoption. Google’s Apollo framework enables transferable black-box optimization for accelerator design, while Pathways coordinates large-scale training across TPU pods with explicit hardware–software co-design \citep{yazdanbakhsh_apollo_2021, barham2022pathwaysasynchronousdistributeddataflow}. Meta’s MTIA accelerator exemplifies a holistic approach that aligns silicon design with specific model families, complemented by fleet-level co-design efforts that tune model parameters to hardware characteristics \citep{coburn_metas_2025, wongpanich_machine_2025}. Commercial tools such as Siemens Simcenter HEEDS apply surrogate models to accelerate engineering workflows, demonstrating the broader transfer of these ideas beyond hyperscale environments \citep{noauthor_simcenter_nodate}.

\subsubsection{Opportunities}

Despite progress, the most pressing bottleneck in generative systems design is verification. While generative models can produce architecture or RTL designs rapidly, evaluating correctness and performance still requires many simulations to account for variability ~\citep{mazurek2023rigorous} or slow formal verification ~\citep{orenes2021autosva, bai2025assertionforge}. This imbalance creates a verification gap where generation far outpaces evaluation. Addressing this gap requires adapting reinforcement learning via verification techniques to hardware design, where fast feedback is scarce and expensive. Performance predictors must therefore evolve from passive estimators into active components of the optimization loop. Models such as Concorde ~\citep{nasr-esfahany_concorde_2025} and NeuSight ~\citep{lee_forecasting_2025} should function as proxy verifiers that provide rapid, approximate feedback to guide generative agents. A hierarchy of verifiers, ranging from coarse but fast approximations to precise but slow simulations, could enable agents to filter most failures quickly while reserving expensive tools for promising candidates. Future work should explore training lightweight verifiers that catch the majority of errors or regressions in milliseconds, enabling iterative self-correction. By closing the loop between generation and verification with such surrogates, the field can move toward agentic design systems that reason about architectural constraints and approach the intuition and efficiency of expert human designers.

\begin{takeawaybox}
\textbf{Performance Prediction and Design Space Exploration}
\begin{itemize}
    \item \textbf{Hybrid analytical-learning performance predictors.} To overcome the prohibitive cost of cycle-accurate simulation, recent systems like Concorde and NeuSight integrate domain-specific analytical bounds with neural networks, enabling rapid, interpretable performance estimation that generalizes better to unseen devices than purely statistical approaches.
    \item \textbf{Generative design space exploration.} Frameworks such as ArchGym and HyperMapper standardize the search for Pareto-optimal architectures using reinforcement learning and Bayesian optimization, facilitating industrial-scale co-design as demonstrated by Google's TPU floorplanning and NVIDIA's arithmetic circuit optimization.
    \item \textbf{Mitigating the verification bottleneck.} A critical disparity exists where generative models produce designs significantly faster than simulators can validate them; addressing this requires the development of active proxy verifiers that provide approximate, millisecond-scale feedback to guide optimization loops before deploying expensive ground-truth verification.
    \item \textbf{Cross-stack patterns.} This section most directly illustrates the feedback loop crisis (C1) and the tacit knowledge problem (C2), with effective responses grounded in hybrid approaches (P1) and matching methods to problem structure (P4). See Section~\ref{sec:challenges_principles}.
\end{itemize}
\end{takeawaybox}

\subsection{Hardware Accelerators and AI Mappings}

The rapid rise of deep neural networks has triggered a fundamental shift in computing, as their computational and memory demands have saturated the capabilities of traditional general-purpose CPUs. This pressure has driven the widespread adoption of hardware accelerators, including GPUs, TPUs, and FPGAs, which provide the massive parallelism and energy efficiency required for modern AI workloads. These accelerators achieve high performance by replacing general-purpose control logic with specialized functional units tailored to core operations such as matrix multiplication and convolution. As a result, accelerator design has become tightly coupled with the structure of AI workloads themselves.

However, achieving high performance on an accelerator depends not only on the hardware but also on how a model is mapped onto it. The AI mapping problem involves translating a high-level neural network into low-level execution decisions that determine graph transformations, operator fusion, memory placement, and kernel scheduling. This mapping space is extremely high dimensional, and suboptimal choices can leave powerful accelerators operating far below peak utilization. Consequently, the field has increasingly turned to AI-driven methods to automate mapping and optimization, enabling systems to explore configurations that outperform manual tuning.

\subsubsection{Datasets and Benchmarks}

Early benchmarking efforts focused on evaluating hardware efficiency through complete model workloads. Benchmarks such as Fathom and the TensorFlow Benchmark Suite introduced representative deep learning workloads spanning vision, speech, and language tasks \citep{Adolf_2016,tensorflow2017benchmarks}. These benchmarks exercised complete training runs on fixed model architectures, reflecting realistic system usage and highlighting hardware throughput on end-to-end tasks. However, because the models were fixed, these benchmarks did not incentivize algorithmic or architectural innovation.

DAWNBench shifted the evaluation paradigm by introducing time-to-accuracy as the primary metric \citep{coleman2017dawnbench}. Rather than fixing the model, it defined target accuracy thresholds and allowed participants to optimize any aspect of the stack to minimize training or inference time. This flexibility enabled innovations such as progressive resizing and quantization but complicated direct hardware comparison due to the diversity of models and techniques employed. At the system level, suites such as SLDB stress model coherence across heterogeneous SoC designs and measure whether generated artifacts maintain intermodule consistency and integration correctness \citep{alvanaki2025sldb}.

MLPerf emerged as an industry-wide standard to balance realism and comparability \citep{reddi2020mlperf}. It combines representative tasks with fixed reference models, accuracy targets, and reproducibility rules across training and inference. MLPerf results have demonstrated rapid progress in accelerator performance and software-hardware co-optimization, while ongoing updates incorporate new workloads and efficiency metrics. Complementing these system-level benchmarks, TenSet introduced a large-scale dataset of operator schedules and performance measurements across hardware backends, enabling supervised training of compiler cost models and dramatically reducing auto-tuning overhead \citep{zheng2021tenset}. By serving as a common training and evaluation resource, TenSet has played a role analogous to ImageNet for learning-based compilation.

\begin{figure}[h!]
    \centering
    \includegraphics[width=0.9\linewidth]{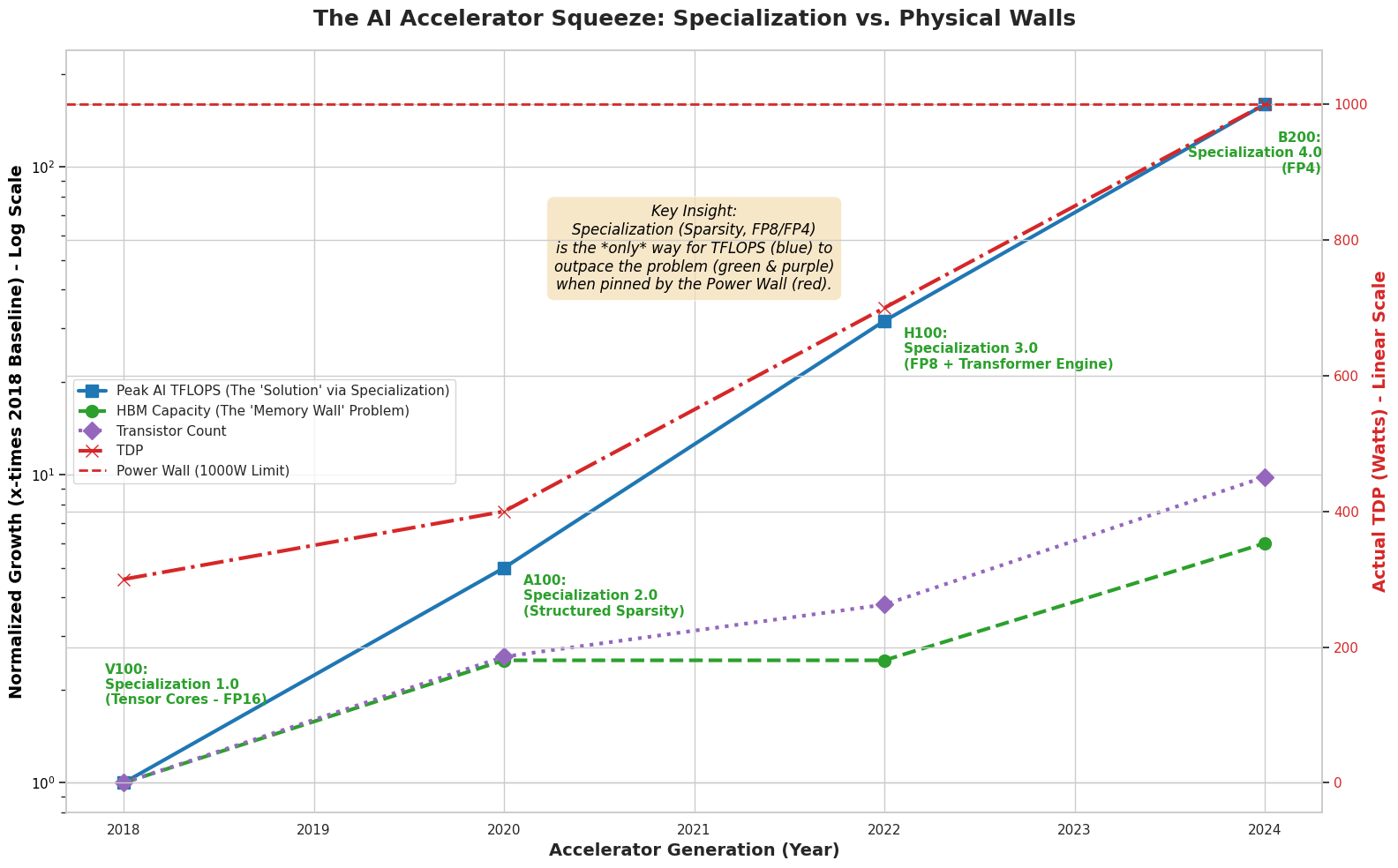}
    \caption{Trends in AI accelerator specialization across recent generations, illustrating how domain-specific features enable compute scaling despite power and memory constraints.}
    \label{fig:hardware_accelerators}
\end{figure}

\subsubsection{Algorithms and Methods}

General-purpose GPUs formed the backbone of deep learning acceleration throughout the 2010s and remain central today, as shown in Figure ~\ref{fig:hardware_accelerators}. Early CUDA-enabled GPUs demonstrated that SIMT architectures matched neural network parallelism well \citep{krizhevsky2012imagenet}. Over time, GPUs incorporated increasingly specialized tensor units. NVIDIA’s Volta architecture introduced Tensor Cores and high-bandwidth memory, while Ampere and Hopper expanded precision formats, sparsity support, and memory systems to deliver substantial gains in training and inference performance \citep{nvidia2017v100,NVIDIA2020A100,NVIDIA2022Hopper}.

Domain-specific ASICs pursued even greater efficiency. Google’s TPU v1 deployed a large systolic array optimized for inference, achieving significant speedups and energy efficiency over CPUs and GPUs \citep{jouppi2017indatacenterperformanceanalysistensor}. Subsequent TPU generations added floating-point support and scalable interconnects, validating the ASIC approach for large-scale training \citep{google2019tpuv3pod}. In parallel, academic accelerators such as Eyeriss demonstrated the importance of dataflow-aware spatial architectures that maximize data reuse across memory hierarchies \citep{chen2016eyeriss}. Processing-in-memory architectures offer another path to efficiency by reducing data movement, where EPIM introduces lightweight epitome operators tailored to PIM constraints, achieving significant area savings while maintaining accuracy through PIM-aware quantization \citep{wang2024epim}. These ideas influenced modern compiler stacks, including TVM, which automate tiling, fusion, and scheduling across diverse backends \citep{chen2018tvm}.

Learning-based mapping and compilation techniques have become increasingly prominent. AutoTVM introduced learned cost models with Bayesian optimization to select efficient schedules \citep{Chen2018AutoTVM}. Ansor and MetaSchedule generalized this approach using template-free search and probabilistic representations scalable across hardware targets \citep{shao2022metaschedule}. At the graph level, TASO framed optimization as a semantics-preserving search over verified rewrites, delivering substantial speedups for complex models \citep{jia2019taso}. Reinforcement learning has also been applied to kernel scheduling and algorithm discovery, as in FlexTensor, while earlier systems demonstrated the feasibility of learned heuristics for compiler flags, throughput prediction, and vectorization \citep{zheng2020flextensor,cummins2017end,haj2020neurovectorizer}.

\subsubsection{Real-World Deployment}

Hardware accelerators and AI mapping techniques have seen widespread deployment across cloud and edge environments. Google’s TPUs have been serving production workloads since 2015, delivering large throughput and energy-efficiency gains for inference and enabling scalable training in TPU pods \citep{jouppi2017indatacenterperformanceanalysistensor,google2019tpuv3pod}. These accelerators are now offered through Google Cloud, demonstrating the viability of specialized hardware at datacenter scale.

Microsoft pursued a different approach with Project Brainwave, deploying FPGA-based neural processing units in Azure datacenters to achieve ultra-low-latency inference without batching. By deploying a configurable accelerator in reconfigurable logic, Brainwave delivered significant latency and throughput improvements while retaining adaptability for evolving models. GPUs remain heavily deployed across cloud providers and supercomputing systems, supporting tasks ranging from large-scale training to inference services. Systems powered by NVIDIA GPUs have enabled training of models with hundreds of billions of parameters, underscoring the continued relevance of general-purpose accelerators with mature software ecosystems.

\subsubsection{Opportunities}

Despite progress, several challenges remain. Current benchmarks prioritize throughput and time-to-accuracy, while real deployments are increasingly constrained by power, energy, and memory-system limits. As models grow, off-package data movement dominates both energy consumption and latency, making memory and interconnect efficiency central concerns. Reliability is another issue, as learned compilers and auto-schedulers can be brittle across software and firmware changes and are difficult to verify formally.

Recent advances in generative AI suggest new directions for accelerator design itself. Methods such as DiffAxE and DiffuSE treat architecture generation as a conditional generative task, sampling accelerator configurations that meet performance and energy targets far more rapidly than traditional search \citep{ghosh2025diffaxediffusiondrivenhardwareaccelerator,Ren2025DiffuSE}. These approaches hint at a future where spatial architectures, memory hierarchies, and dataflow pipelines can be designed with the same speed and flexibility as software, enabling hardware to evolve in step with rapidly changing AI workloads.

\begin{takeawaybox} \textbf{Hardware Accelerators and AI Mappings} \begin{itemize} \item \textbf{Transition toward Memory-Centric Benchmarking.} While early efforts focused on compute throughput \citep{Adolf_2016}, the field has shifted toward end-to-end metrics like time-to-accuracy in MLPerf \citep{reddi2020mlperf} and datasets like TenSet \citep{zheng2021tenset}. These reflect a reality where the "Memory Wall" of off-package data movement and HBM bandwidth frequently dominates performance over raw tensor-op counts. \item \textbf{Bridging the Compiler-Hardware Boundary.} To navigate the high-dimensional mapping space, compilers like TASO \citep{jia2019taso} and MetaSchedule \citep{shao2022metaschedule} leverage Bayesian optimization and verified graph rewrites. These tools prioritize memory-hierarchy-aware optimizations to maximize the utilization of specialized units like systolic arrays \citep{jouppi2017indatacenterperformanceanalysistensor} and Tensor Cores \citep{NVIDIA2022Hopper}. \item \textbf{Generative Synthesis for Rapid Specialization.} Moving beyond heuristic search, emerging frameworks like DiffAxE \citep{ghosh2025diffaxediffusiondrivenhardwareaccelerator} and DiffuSE \citep{Ren2025DiffuSE} treat architecture design as a conditional generative task. This allows for the rapid sampling of spatial architectures and memory hierarchies that are co-optimized for specific model families, evolving hardware at the pace of software. \item \textbf{Cross-stack patterns.} This section illustrates the Feedback loop crisis (C1) where simulation of complex memory systems is slow, and Co-design across boundaries (C4). Successful responses center on Matching methods to problem structure (P4) by aligning tensor graph topology with hardware dataflow, while Building on systems knowledge (P5) regarding locality and bandwidth. See Section~\ref{sec:challenges_principles}. \end{itemize} \end{takeawaybox}
\subsection{Memory Systems and Data Management}

\begin{figure}[h!]
    \centering
    \includegraphics[width=\linewidth]{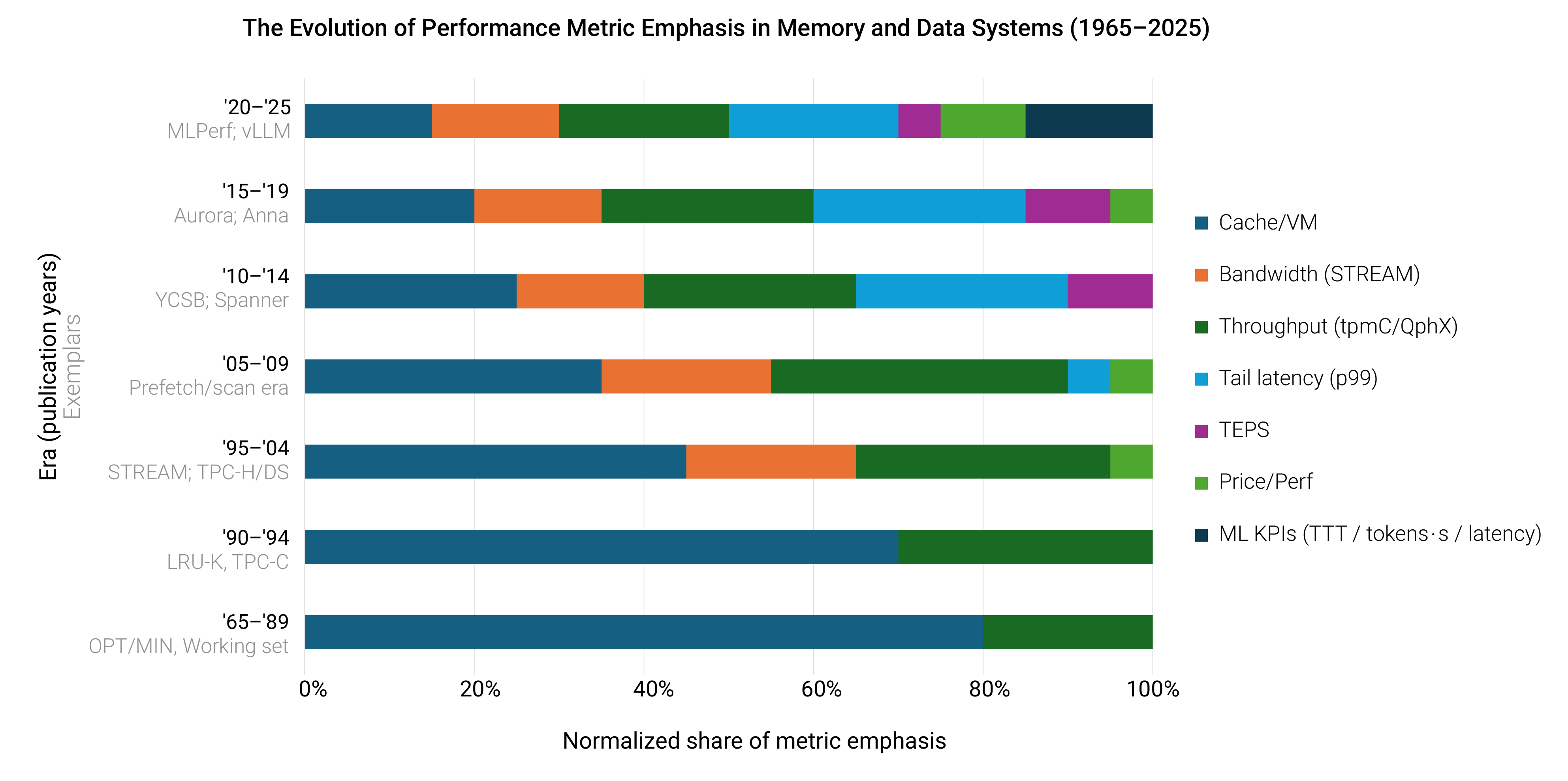}
    \caption{The evolution of performance priorities in memory and data systems from 1965 to 2025, showing how metric families such as cache, bandwidth, tail latency, and ML KPIs wax and wane over time.}
    \label{fig:perf-evolution}
\end{figure}

Memory systems and data management remain central to system performance, but the definition of performance has shifted with workload trends. Early work emphasized locality and working sets, then attention moved to throughput and bandwidth for scan heavy analytics, later to tail latency and coordination in distributed services, and most recently to cost and ML centric KPIs such as time to train and inference throughput. Figure~\ref{fig:perf-evolution} summarizes this trajectory and shows that benchmarks steer system design. Locality remains important, but modern evaluation is multi objective and context bound, with acceptable trade offs determined by workload mix, scale, and service level objectives.

This section is organized around the evidence and mechanisms needed to reason about multi objective, context dependent performance. We first review datasets and benchmarks with an emphasis on trace driven evaluation that replays recorded microarchitectural instruction and memory access streams as well as cluster and service level traces. We then discuss algorithms and methods addressing hot set management across heterogeneous tiers, learned prefetching and input pipeline orchestration, lifetime aware huge page allocation, consistency aware caching and placement, and trace aware scheduling. We close by outlining practical opportunities and challenges in measuring ML KPIs, tail latency, bytes moved across tiers, and cost or energy.

\subsubsection{Datasets and Benchmarks}

Trace driven datasets are well suited to multi objective, context bound evaluation because they replay realistic microarchitectural and cluster behaviors. At the micro level, newer versions of Google Workload Traces provide instruction and memory reference traces from production warehouse scale workloads and are used to study cache, TLB, and memory hierarchy behavior under phase changes and working set shifts \citep{clusterdata:Wilkes2020a}. At the cluster level, the Google cluster usage traces from 2011 and 2019 document jobs, tasks, and machine events and support evaluation in realistic management scenarios that include heterogeneity, over commitment, and heavy tailed resource usage \citep{clusterdata:Wilkes2011,clusterdata:Wilkes2020a,clusterdata:Verma2015,clusterdata:Tirmazi2020}.

Trace analysis guides both understanding and controller design. Early analyses showed bimodal job durations and concentration of resource usage in a few long running tasks \citep{clusterdata:Mishra2010}. Later work emphasized burstiness, configuration diversity, and non stationary load patterns that drive scheduling delay and utilization. These properties make trace based evaluation essential for policies that must manage time varying memory pressure and colocation effects. Architecture reasoning benchmarks such as QuArch probe higher order design knowledge and reveal persistent gaps in advanced architectural reasoning of memory systems \citep{prakash2025quarch}.

There is growing interest in synthetic and generative traces for privacy preserving sharing and stress testing. DoppelGANger uses GANs to synthesize networked time series including cluster request workloads \citep{clusterdata:Lin2020}. Synthetic traces can expand evaluation coverage while protecting privacy, but they raise fidelity concerns, especially about preserving tail events and rare failure modes. Overall, trace driven data supports replay experiments that measure p95 and p99 latency, time to train to a quality target, bytes moved across tiers, and cost or energy per useful unit of work.

\subsubsection{Algorithms and Methods}

Work on memory systems and data management has shifted from single-objective heuristics toward multi-objective, context-aware control driven by realistic traces. Rather than optimizing cache hit rate or throughput in isolation, modern methods reason about tail latency, cost, and ML-centric KPIs under non-stationary, multi-tenant workloads. Learned models are used both as controllers and as evaluation signals, with traces serving as the primary substrate for training, validation, and comparison.

Hot-set management across heterogeneous memory tiers now reasons about value per byte rather than recency alone. In AI pipelines, hotness is highly phase- and tenant-dependent, so learning-based managers extract features such as reuse distance, execution phase, and tenant identity from traces to predict reuse and guide tiering decisions. Prefetching and input-pipeline orchestration similarly coordinate depth, parallelism, and placement across storage, host memory, and accelerator memory. Trace-based studies show that naive always-on prefetching can worsen tail latency under multi-tenant interference, motivating learning-based prefetchers that treat memory streams as sequences and use RNNs or lightweight classifiers to enable phase-aware toggling based on queueing delay and utilization~\citep{hashemi_learning_2018,alcorta2023lightweight}.

Virtual memory and allocation techniques address the large address-space footprints of modern models. Lifetime-aware allocators learn object lifetimes from traces and co-locate objects on huge pages to reduce fragmentation and TLB pressure while preserving flexibility for dynamic workloads~\citep{maas_learning-based_2020}. Learned indexes and ML-augmented data structures use compact models to map keys to positions or existence probabilities, exploiting skew to reduce footprint and latency. Adaptive designs such as ALEX support updates while co-designing models and layout for mixed workloads~\citep{kraska2018case,ding_alex_2020}. Consistency-aware caching and placement further couple admission and eviction with consistency and quorum choices to trade freshness against p95 and p99 latency, with trace-driven evaluation showing that ignoring skew, priority, or failure modes degrades tail behavior. Finally, learning-based controllers and generative models tie these components together: schedulers and placement systems trained on cluster traces predict runtimes and affinities and are evaluated in closed loop~\citep{clusterdata:jajooSLearnNSDI2022,clusterdata:Sliwko2024,clusterdata:Sliwko2025}, while generative trace models such as DoppelGANger enable synthetic workloads for stress testing and privacy-aware sharing, albeit with careful validation required for tail behavior~\citep{clusterdata:Lin2020}.

\subsubsection{Real-World Deployment}

Trace-driven validation is standard practice for transitioning memory-systems research into production, particularly in hyperscale environments where tail latency, energy, and cost are first-class constraints. Large cloud providers routinely rely on production traces to train and validate learned controllers before deployment, ensuring that policies generalize across workload phases, multi-tenant interference, and hardware heterogeneity. A clear example of this is Google’s deployment of a learning-based memory allocation strategy in production services. In the LLAMA framework, supervised models learn object lifetime distributions from real allocation and memory-access traces and use these predictions to co-locate long-lived objects on huge pages. This reduces TLB pressure and memory fragmentation in large C++ services without sacrificing flexibility, demonstrating that ML predictors grounded in realistic traces can safely improve core OS-level memory mechanisms in live systems~\citep{maas_learning-based_2020}.

Real deployments of learned memory presets and tiering policies tend to follow a cautious pattern. Learned controllers are typically embedded within hybrid control loops that include rule-based safety checks, explicit SLO enforcement, and deterministic fallbacks because worst-case behavior (e.g., tail latency spikes) remains operationally critical. For example, in serving and training pipelines that leverage both host and accelerator memory, learned prefetchers and pacing policies are validated on trace populations before being applied, and are gated by conservative thresholds so that rare extremes do not violate service level objectives. Production fleets still monitor systems with traditional performance counters and guard rails, blending learned policies with well-understood heuristics to balance adaptivity and predictability. Operators also use synthesized or generative traces for stress testing and privacy-preserving evaluation, but remain careful about trusting synthetic tails without corroboration from real traces. Synthetic generators, such as GAN-based cluster trace models, expand evaluation coverage and help validate controllers under outlier conditions, but require extensive calibration against ground truth before deployment. In practice, production memory systems blend ML components with rule-based safety checks, deterministic fallback policies, and operational monitoring, reflecting the high cost of regressions and the need for verifiable reliability in production environments.

\subsubsection{Opportunities}

Benchmarks must expand beyond single metric goals to include energy, cost, and ML KPIs such as time to train and tokens per second at a quality target. Reports should include energy per step or per token and power capped runs to reveal behavior under realistic constraints. For learned controllers benchmarks should expose learning overheads, convergence dynamics, and robustness under workload shifts while preserving consistency and fairness.

Generative models act primarily as stressors that amplify system demand. Training pipelines assemble multi petabyte datasets, shard model state across accelerators, and generate bursty traffic. Inference with long contexts increases KV cache footprints and feature store load. Systems research should therefore treat generative workloads as drivers for coordinated placement, paging, caching, and scheduling across tiers.

ML and GenAI are also useful design tools that sit off the critical per access path. Small task specific models perform prediction and RL policies trained on traces can manage control, while generative models synthesize workloads, summarize configurations, and map informal SLOs to concrete parameters. Key challenges are ensuring synthetic traces preserve rare and tail behaviors and building uncertainty aware models with safe fallbacks so that learned controllers are trustworthy by default.

\begin{takeawaybox}
\textbf{Memory Systems and Data Management}
\begin{itemize}
    \item \textbf{Trace-driven evaluation of multi-objective performance.} Benchmarks have shifted from isolated throughput metrics to multi-objective ML KPIs, such as time-to-train and inference tail latency, necessitating trace-driven evaluation using datasets like Google Workload Traces \citep{clusterdata:Wilkes2011, clusterdata:Wilkes2020a}. These traces capture critical non-stationary behaviors and resource burstiness, enabling the design of controllers that balance performance against cost and energy constraints.
    \item \textbf{ML-augmented memory allocation and prefetching.} Modern memory systems replace static heuristics with learned models, such as lifetime-aware allocators that reduce TLB pressure by co-locating long-lived objects on huge pages \citep{maas_learning-based_2020}. Similarly, learning-based prefetchers utilize RNNs to manage input pipeline orchestration and hot-set management, adapting to phase changes and multi-tenant interference where naive policies fail \citep{hashemi_learning_2018}.
    \item \textbf{Hybrid control for safe production deployment.} Production deployments wrap learned controllers in hybrid loops with deterministic guard rails to prevent service level objective (SLO) violations during tail latency spikes. While generative models like DoppelGANger \citep{clusterdata:Lin2020} provide synthetic stress testing, operators prioritize trace-driven validation to ensure reliability across heterogeneous hardware and bursty GenAI workloads.
    \item \textbf{Cross-stack patterns.} This section most directly illustrates co-design across boundaries (C4) and the shift from determinism to dynamism (C5), with effective responses grounded in hybrid approaches (P1) and building on existing systems knowledge (P5). See Section~\ref{sec:challenges_principles}.
\end{itemize}
\end{takeawaybox}
\subsection{LLM Systems and AI Workload Scheduling}

Large language models have transformed AI inference into a complex, stateful workload characterized by autoregressive token generation, variable output lengths, and interactive request patterns. Unlike traditional stateless inference, LLM serving involves a two-phase execution profile consisting of a compute-heavy prefill phase followed by a memory-dominated decoding phase, during which a growing key–value cache must be maintained~\citep{pope2023efficiently}. These characteristics introduce tight coupling between queuing, batching, memory hierarchy management, and scheduling decisions. At the same time, modern applications impose strict latency targets such as time-to-first-token alongside cost and throughput constraints, particularly in chat-based and agentic workflows~\citep{zhang2026agenticcontextengineeringevolving}. As a result, effective LLM serving requires joint optimization across system layers, informed by model-aware algorithmic choices.

This section surveys the datasets, optimization methods, and deployed systems that shape LLM workload scheduling today. A recurring theme is architecture–systems co-design, where model-level levers such as routing, quantization, and speculative execution interact closely with system-level constraints on memory, batching, and resource allocation.

\subsubsection{Datasets and Benchmarks}

Evaluation of LLM serving systems has evolved from synthetic replays toward trace-driven and standardized benchmarks that capture realistic throughput, latency, and cost trade-offs. Production traces play a central role in characterizing arrival processes and burstiness. Public datasets released by cloud providers, such as Azure traces~\citep{cortez2017resource}, and application-driven traces such as BurstGPT~\citep{wang2025burstgpt} provide ground truth for queuing and batching studies under realistic load. These traces reveal highly variable request patterns that stress both schedulers and memory systems.

To emulate interactive behavior, multi-turn conversational corpora such as LMSYS-Chat-1M~\citep{zheng2023lmsys} and ShareGPT capture coupled input and output length distributions characteristic of chat applications. Instruction-tuning datasets~\citep{peng2023instruction} serve as simpler single-turn baselines that isolate model behavior without conversational context. Together, these datasets allow researchers to study how conversational structure influences batching efficiency, latency, and KV-cache growth.

Simulation frameworks complement trace-driven evaluation by enabling controlled design space exploration. System-level predictors such as VIDUR estimate end-to-end performance metrics across configurations without full deployment~\citep{agrawal2024vidur}, while operator-level simulators like ReALLM isolate the impact of architectural changes such as quantization or kernel selection~\citep{leconte2024reallm}. In practice, trace-driven simulation is preferred for scheduling and queuing studies, whereas operator-level frameworks are better suited for kernel- and model-level research. Standardized suites such as MLPerf Inference~\citep{reddi2020mlperf} and LLM-Inference-Bench~\citep{chitty2024llm} enable cross-hardware comparison under fixed service-level objectives, while recent work proposes energy efficiency LLM benchmarking ~\citep{tschand2025mlperfpowerbenchmarkingenergy, saad2025intelligence}, cost-normalized meta-metrics ~\citep{salaria2025meta}, and specialized harnesses for techniques such as speculative decoding to improve experimental rigor~\citep{xia-etal-2024-unlocking}.

\begin{figure}[t]
    \centering
    \includegraphics[width=0.75\linewidth]{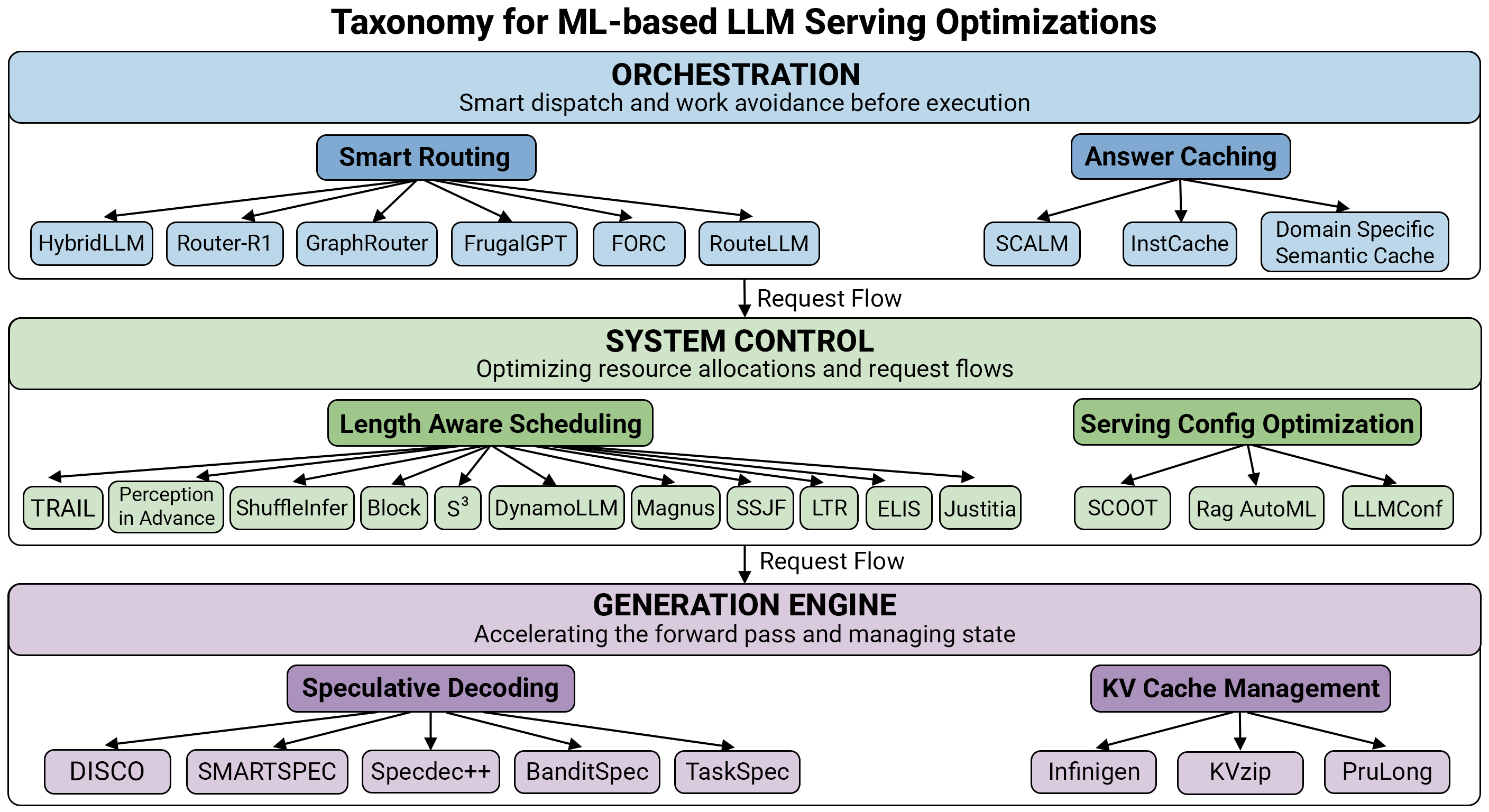}
    \caption{Taxonomy of learning-based optimizations for LLM serving, spanning orchestration, system control, and generation engine mechanisms.}
    \label{fig:llm_scheduling_taxonomy}
\end{figure}

\subsubsection{Algorithms and Methods}

Optimization techniques for LLM serving can be organized into three interacting layers: orchestration, system control, and the generation engine, as illustrated in Figure~\ref{fig:llm_scheduling_taxonomy}. At the orchestration layer, smart routing and answer caching aim to balance quality, latency, and cost. Systems route requests across model cascades or heterogeneous model pools using learned policies based on reinforcement learning, graph neural networks, or cost-aware heuristics~\citep{chen2023frugalgpt,feng2024graphrouter,dai2024cost}. Recent work also explores coordinating multiple small language models to achieve higher accuracy through complementary specialization and test-time scaling~\citep{wang2025slm}. Answer caching reduces redundant computation by serving previously generated outputs for identical or semantically similar prompts, with improvements driven by embedding-based similarity and predictive caching models~\citep{bang2023gptcache,zou2024instcache}. For agentic workflows, plan caching extends this principle to multi-step reasoning traces, enabling cost-efficient reuse of intermediate plans~\citep{zhang2025cost}.

At the system control layer, learned scheduling and configuration aim to improve batching and resource utilization. Predicting output length is particularly valuable because it enables schedulers to group compatible requests and reduce head-of-line blocking. Methods range from self-prediction by the LLM~\citep{zheng2023response} and hidden-state analysis~\citep{shahout2024don} to learning-to-rank approaches and regression models trained on logs~\citep{fu2024efficient,akhauri2025performance}. These predictors are typically compact models such as BERT variants or MLPs, and they deliver substantial throughput and latency improvements over first-come-first-served baselines, although direct comparison across studies is difficult due to differing assumptions and platforms. Beyond scheduling, configuration optimization systems such as SCOOT and LLMConf tune complex serving hyperparameters using surrogate models to navigate large configuration spaces efficiently~\citep{cheng2025scoot,he2025llmconf}.

The generation engine layer focuses on accelerating token generation and managing memory bottlenecks. Speculative decoding uses smaller draft models to propose candidate tokens that are then verified by the full model, reducing effective compute per output token~\citep{leviathan2023fast}. Learned policies predict optimal draft lengths or select draft models dynamically using bandits or lightweight predictors~\citep{mamou2024dynamic,hou2025banditspec,ge2025automatictaskdetectionheterogeneous}. KV-cache management addresses the dominant memory cost of long-context inference through learned compression schemes, such as codebooks or vector quantization, and through eviction or attention-gating policies that selectively retain useful context~\citep{kim2024lexico,li2025commvq,xiao2024duoattention,lee2024infinigen}.

\subsubsection{Real-World Deployment}

In practice, the LLM serving ecosystem has converged around a small set of high-performance inference engines. Systems such as vLLM, which introduced paged attention, as well as TGI and TensorRT-LLM, standardize continuous batching, memory-efficient attention, and optimized kernels. These engines are commonly integrated with serving platforms such as Ray Serve and KServe, which provide autoscaling, routing, and fault tolerance.

Recent deployments emphasize architectural refinements that address the interference between prefill and decode phases. Disaggregated serving architectures, such as Splitwise~\citep{patel2024splitwise} and DistServe~\citep{zhong2024distserve}, decouple these stages onto separate hardware instances to optimize for their distinct compute and memory-bandwidth profiles. This separation allows for independent scaling and prevents high-latency prefill bursts from stalling ongoing decoding streams. These systems are often evaluated using simulators like VIDUR~\citep{agrawal2024vidur} to navigate the trade-offs between throughput and time-to-first-token. In parallel, specialized frameworks now support multi-adapter batching and chunked prefill to further increase utilization in multi-tenant environments.

\subsubsection{Opportunities}

As LLMs evolve toward longer contexts, richer reasoning, and alternative generation paradigms, system bottlenecks continue to shift. Scaling test-time compute for reasoning-heavy workloads requires schedulers that can allocate variable compute budgets and manage extended thinking traces. Training and serving are also becoming more tightly coupled, as reinforcement learning from human feedback and online fine-tuning demand systems that interleave generation rollouts with weight updates efficiently.

Emerging non-autoregressive models such as diffusion-based LLMs challenge existing assumptions about KV caching and batching, requiring new memory abstractions and scheduling policies. More broadly, compositional AI systems that integrate LLMs with symbolic reasoning, probabilistic inference, or collections of specialized models introduce heterogeneous kernels, irregular memory access patterns, and complex data dependencies that strain current serving stacks~\citep{wan2025compositional,wang2025slm}. These trends create opportunities for deeper co-design across model architectures, scheduling algorithms, quantization strategies, and heterogeneous accelerator pools coordinated by dynamic, policy-driven routing.

\begin{takeawaybox}
\textbf{LLM Systems and AI Workload Scheduling}
\begin{itemize}
\item Stateful inference and memory-compute coupling. LLM serving requires managing the distinct resource demands of compute-intensive prefill and memory-dominated decoding phases. This has led to the adoption of PagedAttention ~\cite{kwon2023efficient} in vLLM and disaggregated serving architectures like Splitwise~\citep{patel2024splitwise} and DistServe~\citep{zhong2024distserve}, which physically separate execution stages to minimize latency interference and optimize KV-cache management.
\item \textbf{Hierarchical optimization across system layers.} Effective scheduling integrates orchestration strategies like model cascading~\citep{chen2023frugalgpt} with system-level controls such as learned output length prediction for efficient batching, while generation engines employ speculative decoding and KV-cache compression to alleviate memory bottlenecks.
    \item \textbf{Trace-driven evaluation and emerging complexity.} While current benchmarking relies on production traces like BurstGPT~\citep{wang2025burstgpt} to capture burstiness and conversational structure, future schedulers must adapt to variable test-time compute for reasoning tasks and the irregular memory access patterns of compositional AI systems~\citep{wan2025compositional}.
    \item \textbf{Cross-stack patterns.} This section most directly illustrates co-design across boundaries (C4) and the shift from determinism to dynamism (C5), with effective responses grounded in separating concerns by role (P3) and matching methods to problem structure (P4). See Section~\ref{sec:challenges_principles}.
\end{itemize}
\end{takeawaybox}

\section{Generative AI for Chip Design}
\label{sec:chipdesign}

\subsection{RTL Design and Logic Synthesis}

\begin{figure}[t]
    \centering
    \includegraphics[width=1\linewidth]{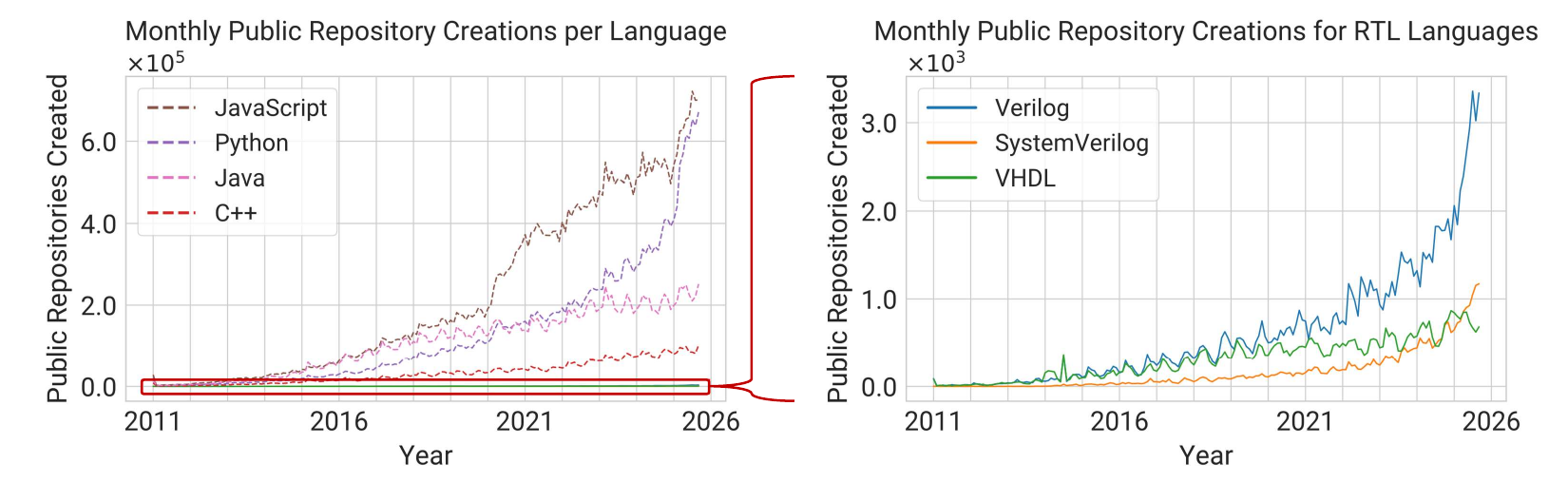}
    \caption{Monthly public GitHub repository creations for RTL compared to non-RTL languages, illustrating the severity of data scarcity.}
    \label{fig:rtl-github-stats}
\end{figure}

Register-Transfer Level design forms the bridge between high-level architecture and physical silicon, but it has become a major bottleneck in modern SoC development~\citep{foster_wilson_report}. While generative AI has rapidly transformed software engineering, RTL design faces a distinct set of challenges. RTL descriptions are inherently parallel, timing-sensitive, and tightly constrained by synthesis and verification requirements. At the same time, the domain suffers from severe data scarcity. As shown in Figure~\ref{fig:rtl-github-stats}, public Verilog repositories account for less than 0.5\% of the volume of mainstream programming languages such as Python or JavaScript, creating a data wall that limits the effectiveness of large pretrained models.

\subsubsection{Datasets and Benchmarks}

Evaluation in RTL generation has evolved from simple syntax checking toward increasingly realistic and system-level benchmarks. Early datasets such as VerilogEval~\citep{liu2023verilogevalevaluatinglargelanguage} and RTLLM~\citep{lu2023rtllmopensourcebenchmarkdesign} focused on single-module completion tasks, including adders and finite state machines. These benchmarks primarily assessed syntactic correctness and basic functional validity, providing a first measure of model capability but limited insight into scalability.

More recent benchmarks address complexity and scale. RealBench~\citep{jin2025realbenchbenchmarkingveriloggeneration} and RTL-Repo~\citep{allam2024rtl} evaluate repository-level generation and composition, requiring models to reason across multiple modules and files. ChipGPTV~\citep{chang2024naturallanguageenoughbenchmarking} extends the input modality by incorporating visual block diagrams, while TuRTLe~\citep{garciagasulla2025turtleunifiedevaluationllms} unifies multiple benchmarks to provide a more holistic evaluation framework. Beyond design generation, specialized benchmarks target downstream tasks such as assertion generation and verification, testbench synthesis, and debugging, reflecting the reality that RTL design is inseparable from verification and iteration~\citep{fang2024assertllmgeneratingevaluatinghardware,SomashekaraMurthy2025VerifLLMBench,wang2025veridebugunifiedllmverilog}.

\subsubsection{Algorithms and Methods}

To address data scarcity and strict physical constraints, algorithmic approaches for RTL generation have specialized along several complementary directions. One prominent trend integrates EDA tools directly into the generation loop. Systems such as AutoChip~\citep{thakur2024autochipautomatinghdlgeneration} and AIvril~\citep{islam2024aivrilaidrivenrtlgeneration} use simulator and compiler feedback to iteratively repair generated RTL, substantially improving functional correctness at the cost of increased latency. VerilogCoder~\citep{ho2025verilogcoderautonomousverilogcoding} extends this approach by incorporating waveform tracing to guide refinement.

Another line of work focuses on data-centric improvements that enrich the limited training signal. Approaches such as MasterRTL~\citep{fang2023masterrtl} and MEV-LLM~\citep{Nadimi_2024} augment RTL with semantic graphs or complexity annotations. Retrieval-augmented generation techniques ground models in trusted IP libraries or design examples, reducing hallucinations and improving robustness, as demonstrated by systems like AutoVCoder~\citep{gao2024autovcodersystematicframeworkautomated} and HDLCoRe~\citep{ping2025hdlcoretrainingfreeframeworkmitigating}.

Agent-based orchestration further decomposes the design process into specialized roles. Multi-agent systems assign coding, verification, and critique tasks to distinct agents that interact through tool feedback, mimicking human design teams. Nexus~\citep{sami2025nexuslightweightscalablemultiagent} provides a hierarchical framework for such coordination, while AIvril2~\citep{islam2024edaawarertlgenerationlarge} demonstrates iterative refinement through agent collaboration. Finally, explicit reasoning and control mechanisms have been introduced to improve logical consistency. ScaleRTL~\citep{deng2025scalertlscalingllmsreasoning} and CodeV-R1~\citep{zhu2025qimengcodevr1reasoningenhancedveriloggeneration} employ chain-of-thought reasoning and reinforcement learning with verifier-derived rewards, EARL~\citep{shi2025earl} uses entropy-aware RL alignment to improve generation reliability, while QiMeng-SALV~\citep{zhang2025qimengsalvsignalawarelearningverilog} optimizes signal-aware preference objectives rather than binary pass or fail feedback.

\subsubsection{Real-World Deployment}

Compared to software generation, deployment of generative models for RTL design remains cautious. Most systems are currently used as assistive tools that accelerate early design exploration, debugging, or verification rather than as fully autonomous generators. Integration with existing EDA flows is critical, and models are typically deployed behind strict validation pipelines that include simulation, synthesis checks, and human review. Early industrial adoption focuses on reducing iteration time for common design patterns and assisting with verification collateral, reflecting the high cost of errors in silicon.

\subsubsection{Opportunities}

Several opportunities remain for advancing generative AI in RTL design. Post-training and alignment methods are needed to move beyond supervised fine-tuning and align models with physical and timing constraints using execution feedback and verifier-guided optimization. Synthetic data generation through procedural specification sampling and back-translation offers a path to bypass the data wall by creating large, correctly labeled curricula~\citep{liu2025craftrtlhighqualitysyntheticdata}. Incorporating power, performance, and area objectives directly into the generation loss would enable PPA-aware design trade-offs rather than purely functional correctness.

Scaling to repository-level design is another challenge, as real engineering workflows rely on incremental changes and engineering change orders. Retrieval mechanisms specialized for module hierarchies and dependency graphs will be required to support safe updates in large codebases. Finally, multimodal reasoning that integrates textual specifications with block diagrams and waveforms could better align generative models with human RTL workflows, closing the gap between high-level intent and low-level implementation.

\begin{takeawaybox}
\textbf{RTL Design and Logic Synthesis}
\begin{itemize}
    \item \textbf{Overcoming data scarcity through repository-scale evaluation.} With public Verilog repositories constituting less than 0.5\% of mainstream language volumes, the field is shifting from simple syntax-checking benchmarks like VerilogEval to complex, repository-level evaluations like RealBench and RTL-Repo that assess a model's ability to reason across module hierarchies and dependencies.
    \item \textbf{EDA-integrated feedback loops and agentic orchestration.} To satisfy strict physical and functional constraints, systems such as AutoChip and AIvril integrate simulator and compiler feedback for iterative code repair, while multi-agent frameworks like Nexus decompose the design process into specialized coding and verification roles to mimic human engineering workflows.
    \item \textbf{PPA-aware alignment and synthetic data generation.} Future research opportunities focus on incorporating power, performance, and area (PPA) objectives directly into generation loss functions, alongside leveraging synthetic data pipelines like CraftRTL to bypass the data wall and align models with multimodal inputs such as block diagrams and waveforms.
    \item \textbf{Cross-stack patterns.} This section most directly illustrates the feedback loop crisis (C1) and the tacit knowledge problem (C2), with effective responses grounded in continuous feedback (P2) and building on existing systems knowledge (P5). See Section~\ref{sec:challenges_principles}.
\end{itemize}
\end{takeawaybox}

\subsection{Physical Design and Layout}

Physical design, particularly chip floorplanning, is one of the most computationally intensive stages of chip design because of the vast design space and complex trade-offs among performance, power, and area under congestion and density constraints. Traditional analytic and stochastic approaches have long dominated this stage, but the rapid scaling of modern chips has strained their effectiveness. Recent agentic demonstrations suggest a new paradigm for layout optimization, yet applying generative AI remains challenging due to sparse public data, enormous design spaces, and limited simulation fidelity. This subsection surveys the datasets, algorithms, deployments, and open challenges that define the transition toward agentic and generative physical design workflows.

\subsubsection{Datasets and Benchmarks}

Half-Perimeter Wirelength remains the most widely used surrogate objective in placement research because it is fast to compute and correlates reasonably with downstream outcomes, enabling controlled experimentation and cross-paper comparison \citep{cheng_replace_2019,lin_dreamplace_2021,zhu_nonsmooth_2015}. However, HPWL neglects congestion, routing detours, power integrity, and timing closure, so minimizing HPWL alone does not guarantee manufacturability or performance. Despite these limitations, HPWL persists as a practical baseline while more PPA-oriented evaluations gain traction.

Classical benchmark suites established empirical continuity for placement research. ISPD-2005 standardized HPWL-based evaluation for post-90\,nm designs \citep{nam_ispd2005}. ICCAD-2012 introduced routability-aware evaluation with realistic blockage constraints, and ICCAD-2019 scaled designs further while incorporating timing and power data to enforce joint PPA optimization \citep{jiang_iccad2012_opening_2012,schlichtmann_iccad2019_overview_2019}. These contests remain central reference points for evaluating analytical, heuristic, and learning-based approaches \citep{cheng_replace_2019,lin_dreamplace_2021}.

Agentic methods motivated new AI-specific testbeds. A2Perf evaluates generalization and transfer of learned policies \citep{uchendu2024a2perf}, while ChiPBench measures end-to-end PPA after full physical design closure rather than relying solely on placement proxies \citep{yu2026chipbenchnextstepbenchmarkevaluating}. Open initiatives such as MacroPlacement emphasize reproducibility through public baselines and standardized scripts, complementing classical benchmarks with robustness and transfer-focused evaluation \citep{tilos_macroplacement,cheng_assessment_2023,markov_reevaluating_2024,mirhoseini_graph_2021}.

To meet the data requirements of modern learning methods, large datasets have emerged. FloorSet provides roughly one million generated floorplans under uniform constraints for representation learning \citep{floorset_arxiv_2024}. CircuitNet~2.0 supplies diverse industrial data, including congestion maps and post-routing artifacts, enabling models to learn distributions of feasible layouts rather than optimizing isolated instances \citep{circuitnet20_iclr_2024}.

\subsubsection{Algorithms and Methods}

\begin{figure}[t]
    \centering
    \includegraphics[width=1.0\linewidth]{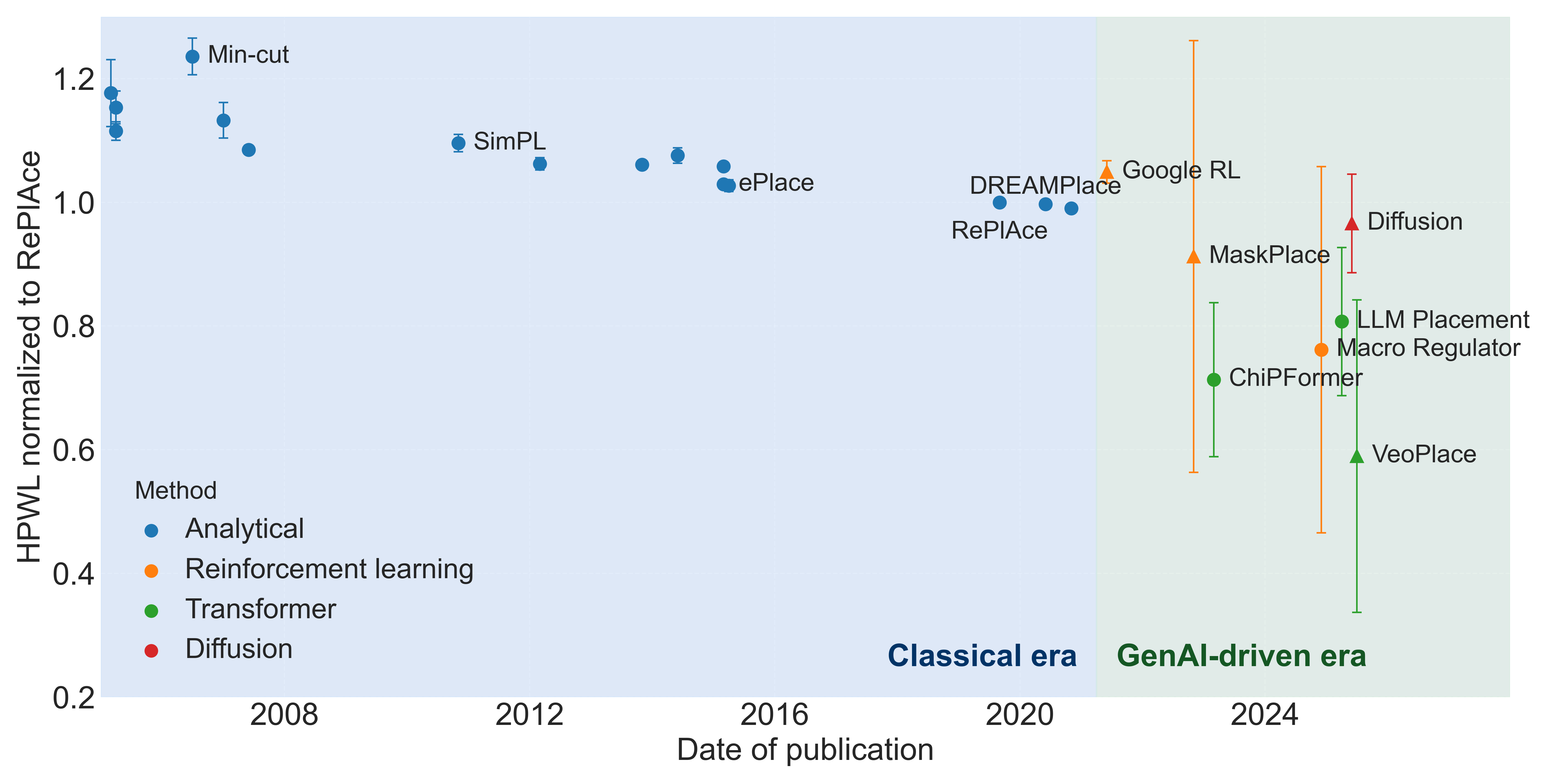}
    \caption{Evolution of placement algorithm performance over time, illustrating the transition from classical optimization to GenAI-driven approaches. Performance is measured by Half-Perimeter Wirelength normalized to the analytical baseline RePlAce \citep{zhu_nonsmooth_2015,lu_eplace_2015,cheng_replace_2019,lin_dreamplace_2021,gu_dreamplace_2020,mirhoseini_graph_2021,lai_maskplace_2022,lai2023chipformer,macro-regulator,yao2025evolutionoptimizationalgorithmsglobal,lee_chip_2025,uchendu2025see}.}
    \label{fig:hpwl_timeline}
\end{figure}

Physical design has a long algorithmic history, with recent work highlighting a shift from classical optimization to agentic and generative frameworks, as shown in Figure ~\ref{fig:hpwl_timeline}. Early methods combined partitioning and stochastic search, including min-cut floorplanning and simulated annealing \citep{roy_min-cut_2006,lichen_efficient_2012}. Deterministic analytical placers later became dominant due to their robustness and convergence properties. SimPL established flat force-directed placement \citep{kim_simpl_2010}, ePlace introduced electrostatic density modeling \citep{lu_eplace_2015}, and RePlAce refined local density functions to become a widely adopted analytical baseline \citep{cheng_replace_2019}. DREAMPlace reformulated this pipeline as a differentiable, GPU-accelerated system, bridging analytical optimization and data-driven learning \citep{lin_dreamplace_2021,gu_dreamplace_2020}.

Reinforcement learning reframed placement as a sequential decision problem over netlist graphs. The graph-based placement work by Mirhoseini et al.\ demonstrated that RL agents could produce competitive macro placements with strong PPA in hours rather than weeks \citep{mirhoseini_graph_2021}. This result sparked debate over reproducibility and generalization, as early implementations relied on proprietary flows \citep{markov_reevaluating_2024}. In response, open frameworks such as Circuit Training were released to improve transparency and scalability \citep{cheng_assessment_2023}. Subsequent work explored hybrid formulations where RL acts as an optimizer or regulator rather than a full placer, combining analytical stability with learned adaptability \citep{agnesina2022parameter,macro-regulator}.

Generative models have further expanded the design space. Transformer-based approaches capture long-range dependencies in netlists and support transfer across designs \citep{lai2023chipformer}. Multimodal methods incorporate visual and textual priors to reason about layout constraints \citep{uchendu2025see}. Diffusion models generate physically plausible layout heatmaps and synthetic datasets, addressing data scarcity and enabling zero-shot placement on unseen circuits \citep{wu_dali-pd_2025,lee_chip_2025}. Other work uses LLMs to suggest algorithmic modifications or evolutionary improvements to existing placers, shifting the role of AI toward design evolution rather than direct placement \citep{yao2025evolutionoptimizationalgorithmsglobal}.

\subsubsection{Real-World Deployment}

Agentic placement has reached production in major industrial flows. Google’s RL-based macro placement reduced planning time for TPU blocks from weeks to hours \citep{mirhoseini_graph_2021,mirhoseini_tpu_nature_2024}. NVIDIA extended similar ideas to large GPU and SoC designs by combining RL-guided macro decisions with differentiable analytical refinement \citep{agnesina2023autodmp}. Commercial EDA tools such as Synopsys DSO.ai and Cadence Cerebrus operationalized learning across the physical design stack, reporting substantial gains in design-space exploration throughput and reduced engineering iteration \citep{synopsys_dsoai_launch_2020,cadence_cerebrus_press_2022}.

Open-source efforts, notably OpenROAD and the Google–SkyWater collaborations, aim to make RTL-to-GDSII flows autonomous, reproducible, and accessible \citep{ajayi2019openroad,openroad_docs_tapeouts}. These initiatives have broadened access to industrial-grade design flows and enabled more transparent evaluation of learning-based methods.

\subsubsection{Opportunities}

Despite rapid progress, many models still optimize proxies rather than true downstream objectives. Improvements in HPWL do not guarantee timing closure, power integrity, or robustness across corners and modes. Advancing beyond copilot tools will require conditioning models on downstream constraints and multi-stage interactions that reflect real design viability.

A promising direction treats physical design as a multimodal problem. Netlists, geometries, timing signatures, power maps, and textual specifications represent complementary views of the same artifact. Models that integrate these signals can begin to reason about physical constraints rather than relying on proxies. Hybrid workflows that combine analytical solvers, learned regulators, and human oversight appear most practical in the near term, balancing robustness with adaptability. Ultimately, the goal is not incremental improvement on HPWL leaderboards, but AI systems that expose trade-offs, clarify design intent, and enable shared reasoning between engineers and models across the full physical design flow.

\begin{takeawaybox}
\textbf{Physical Design and Layout}
\begin{itemize}
    \item \textbf{Transition to learning-based placement algorithms.} While deterministic analytical solvers like RePlAce and the differentiable DREAMPlace remain robust baselines, the field is shifting toward reinforcement learning and generative frameworks, including graph-based RL agents for macro placement and Transformer-based models like ChipFormer that capture long-range netlist dependencies.
    \item \textbf{Limitations of proxy objectives in benchmarking.} Although Half-Perimeter Wirelength (HPWL) serves as a standard surrogate metric for benchmarks like ISPD and ICCAD, it fails to guarantee manufacturability or timing closure; this has spurred the development of end-to-end PPA evaluation suites like ChiPBench and large-scale datasets like CircuitNet~2.0 to support robust representation learning.
    \item \textbf{Industrial deployment and multimodal opportunities.} Agentic placement has achieved production status in flows at Google, NVIDIA, and via tools like Synopsys DSO.ai, yet future advancements require moving beyond single-objective optimization toward multimodal systems that integrate netlists, geometric constraints, and timing signatures to address complex physical design trade-offs.
    \item \textbf{Cross-stack patterns.} This section most directly illustrates trust and validation (C3) and co-design across boundaries (C4), with effective responses grounded in hybrid approaches (P1) and matching methods to problem structure (P4). See Section~\ref{sec:challenges_principles}.
\end{itemize}
\end{takeawaybox}

\subsection{Hardware Verification and Advanced Chip Design}

Hardware verification ensures that hardware implements its intended specification and typically dominates engineering effort and cost in chip development. With the rise of machine learning, verification workflows are being reconceived as pipelines where learned predictors, generative models, and symbolic engines cooperate. In this subsection we survey the datasets and benchmarks that measure progress, the algorithmic methods that integrate ML into formal and simulation flows, the early industrial deployments that show promise, and the remaining opportunities for making ML driven verification reliable and auditable.

\begin{figure}[t!]
    \centering
    \includegraphics[width=\textwidth]{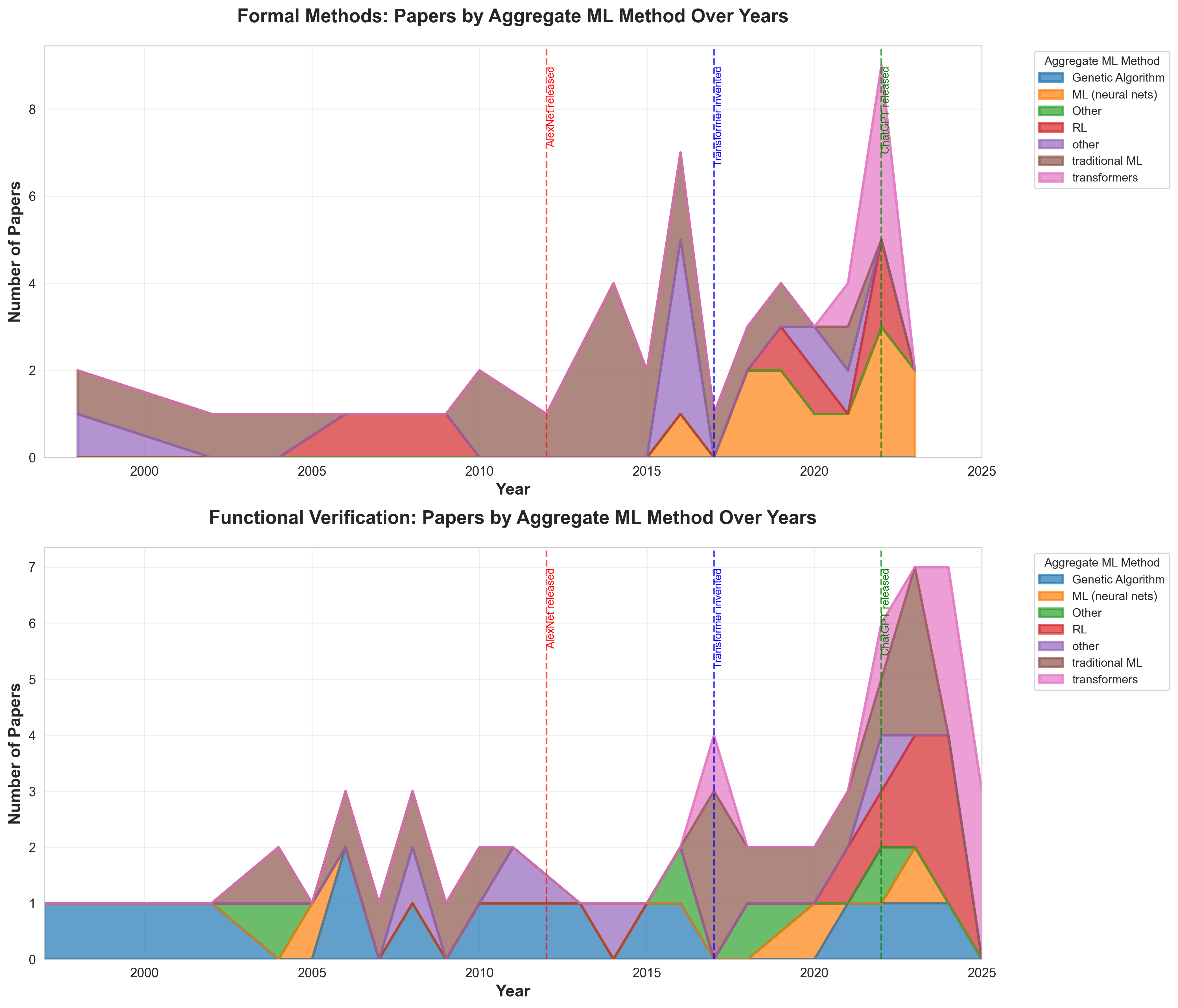}
    \caption{Survey of ML in hardware verification papers by ML method across formal methods and functional verification. Colors encode aggregate ML method.}
    \label{fig:verification-over-time}
\end{figure}

The overview figure \ref{fig:verification-over-time} situates recent papers by method and by verification domain to show how literature has trended over time and to emphasize the growing role of generative models alongside traditional ML techniques.

\subsubsection{Datasets and Benchmarks}

Benchmarks for verification now span component generation, assertion synthesis, and system level integration, each targeting different aspects of correctness and reasoning. Component level datasets evaluate whether models can produce syntactically valid and functionally correct RTL from descriptions and test benches, measuring compilation success and equivalence under simulation \citep{thakur2022benchmarkinglargelanguagemodels}. These datasets typically report metrics for parsing, compilation, and pass rate under unit tests, which surfaces common failure modes like incomplete state machines or incorrect handshakes. The results show that code emission by itself is necessary but not sufficient for verification grade artifacts.

Assertion generation benchmarks and frameworks such as AssertLLM and FVEval test a model’s ability to translate informal intent into SystemVerilog assertions and to reason about properties against RTL code \citep{fang2024assertllmgeneratingevaluatinghardware,kang2024fvevalunderstandinglanguagemodel}. These evaluations move beyond surface correctness by measuring whether generated assertions actually capture intended constraints and whether they help the formal engines converge. 

\subsubsection{Algorithms and Methods}

Algorithmic work in verification divides into two complementary streams. The first stream embeds learned heuristics inside symbolic engines, where models accelerate SAT solving, model checking, and equivalence checking by guiding branching, abstraction, or search policies. Graph neural networks and reinforcement learners have been shown to prune search spaces and to prioritize promising proof paths, improving solver throughput on many industrial instances. Supervised classifiers and sequence models help with equivalence and invariant detection by proposing candidate mappings that symbolic tools then validate.

The second stream leverages large language models to perform higher level translation and synthesis tasks. LLMs are applied to specification to formal translation, testbench scaffolding, assertion synthesis, and to orchestrating verification campaigns \citep{shih2025flagformalllmassistedsva}. Hybrid pipelines pair LLM proposals with symbolic feedback loops so that generated assertions, testbenches, and stimuli are iteratively validated by model checkers or simulators. This combination uses the generative capacity of LLMs to reduce manual effort while relying on formal engines to preserve soundness.

\subsubsection{Real-World Deployment}

Industry is beginning to embed generative and learned components into established verification toolchains while keeping symbolic engines as the source of truth. Cadence and Synopsys have announced integrated assistants that consume documentation and verification data to help with debug, regression triage, and assertion creation \citep{Cadence2022Verisium,synopsys_copilot_generative_ai}. These assistants generate candidate SVAs, testbench fragments, and scripts that are then run under existing simulators and formal tools. Early deployments emphasize a reviewer in the loop model where humans vet and sign off on machine suggested artifacts.

Open source efforts also contribute to deployment readiness by providing reproducible flows and data for evaluation. Projects that publish validation harnesses and reproducible experiments allow researchers to test LLM augmented pipelines on realistic RTL and coverage targets. These public flows make it practical to measure the downstream effect of generated artifacts on coverage closure and formal convergence, and they help surface engineering concerns such as brittleness across toolchain versions.

\subsubsection{Opportunities}

The most urgent research needs center on reliability, calibrated uncertainty, and end to end evaluation. Verification demands guarantees, so hybrid systems must expose confidence measures and produce artifacts that symbolic engines can validate automatically. Work on uncertainty quantification and on techniques that reduce hallucinated or unsafe assertions will be essential for industry adoption. Benchmarks must measure semantic correctness and downstream impact, reporting not only compilability but also assertion coverage, counterexample usefulness, and change in convergence time for formal checks.

Multimodal and domain adapted models offer another promising avenue. Models that jointly consume specs, RTL, waveforms, and coverage reports could reason about intent and produce assertions and tests that target real failure modes. Finally, agentic verification that schedules simulator campaigns, selects heuristics, and invokes formal checks promises major productivity gains, but this will require careful audit trails, rollback mechanisms, and integration patterns that preserve traceability and human oversight. Progress on these fronts could shift ML from a suggestive tool to a trusted partner in the most costly phase of chip development.

\begin{takeawaybox}
\textbf{Verification and Advanced Chip Design}
\begin{itemize}
    \item \textbf{Evolution of verification-specific benchmarks.} Recent frameworks like AssertLLM, FVEval, and SLDB move beyond simple compilation metrics to evaluate semantic correctness, measuring whether generated assertions capture design intent and assist formal engines in convergence rather than just passing syntax checks \citep{fang2024assertllmgeneratingevaluatinghardware,kang2024fvevalunderstandinglanguagemodel,alvanaki2025sldb}.
    \item \textbf{Hybrid neuro-symbolic verification pipelines.} Methodologies increasingly pair the generative capabilities of LLMs for assertion and testbench synthesis with symbolic feedback loops, ensuring that machine-suggested artifacts are rigorously validated by model checkers to preserve soundness while reducing manual engineering effort \citep{shih2025flagformalllmassistedsva}.
    \item \textbf{Reliability and agentic verification challenges.} To transition from suggestive tools to trusted partners, future systems require calibrated uncertainty quantification to mitigate hallucinations and multimodal models that ingest waveforms and specifications, enabling agentic workflows that can autonomously schedule verification campaigns with necessary audit trails.
    \item \textbf{Cross-stack patterns.} This section most directly illustrates trust and validation (C3), with effective responses grounded in hybrid approaches (P1) and separating concerns by role (P3). See Section~\ref{sec:challenges_principles}.
    \end{itemize}
\end{takeawaybox}

\section{The Road Ahead: Toward Systematic AI-for-Systems Engineering}
\label{sec:open_questions}

The preceding sections examined how generative AI is being applied across eleven areas of the computing stack, from code generation and performance engineering through hardware architecture to RTL design, physical layout, and verification. Each section identified layer-specific opportunities and limitations. But the cross-stack view reveals something more: there are research questions that no single layer can answer on its own, because they arise from the interaction between challenges and principles that span the entire stack.

The five challenges and five design principles introduced in Section~\ref{sec:challenges_principles} provide the vocabulary for framing these questions. Where the body sections showed how individual communities are responding to specific instantiations of these challenges, the questions below ask what it would take to resolve them structurally, across layers, rather than one domain at a time. Four open research questions emerge from this cross-stack analysis.

\subsection{Cross-Stack Research Questions}

\paragraph{Open question 1: How do we build evaluation that closes the loop across layers?}
The feedback loop crisis (C1) and the principle of continuous feedback (P2) together highlight a major gap in evaluation infrastructure. Many subfields now face benchmark lifecycle compression, where models quickly saturate static tasks and progress shifts to harder, more realistic settings \citep{merrill2026terminalbenchbenchmarkingagentshard}. A cross-stack opportunity is to build benchmarks that measure not just one artifact, but the end-to-end loop that produced it. In software, this means tasks that require persistent improvement under evolving repositories and dependencies \citep{jimenez2023swebench}. In GPU kernels and LLM serving, it suggests trace-driven suites that connect micro-level kernel decisions to application-level metrics like time-to-train, tokens per second, and tail latency under realistic burstiness \citep{ouyang2025kernelbench,ye2025flashinferefficientcustomizableattention,wang2025burstgpt}. In EDA, it means moving beyond placement proxies toward end-to-end closure metrics, as emphasized by ChiPBench, and toward datasets that capture search trajectories rather than only static design and label pairs \citep{yu2026chipbenchnextstepbenchmarkevaluating,krishnan_archgym_2023}. The broader research question is how to create shared, cross-layer evaluation harnesses where improvements at one layer can be reliably attributed to user-facing outcomes.

\paragraph{Open question 2: What is the right architecture for verified generation at scale?}
Trust and validation (C3) combined with the principles of hybrid approaches (P1) and role separation (P3) point toward a structural question that remains open. The survey suggests a recurring asymmetry between fast generators and slow verifiers, and the solution is unlikely to be a single verifier. Instead, we need hierarchies of proxy verifiers that trade fidelity for speed in a controlled way. In hardware DSE, this is already motivating the view that predictors like Concorde and NeuSight should act as reward functions for generative agents, not just as passive estimators \citep{nasr-esfahany_concorde_2025,lee_forecasting_2025}. In RTL, tool-in-the-loop repair points toward layered checking that starts with syntax, moves through simulation, and escalates to formal methods when needed \citep{thakur2024autochipautomatinghdlgeneration,fang2024assertllmgeneratingevaluatinghardware}. In kernels, robust-kbench reflects the same direction by expanding beyond single-run correctness and into systematic stress testing \citep{lange2025robustkbench}. A unifying open problem is to formalize interfaces between generators and verifiers, including contracts, uncertainty estimates, and escalation policies, so that verification becomes a scalable component of the loop rather than a bottleneck that blocks it.

\paragraph{Open question 3: How can systems learn under churn and still generalize?}
The shift from determinism to dynamism (C5) and the principle of building on existing systems knowledge (P5) raise the question of what happens when the environment itself keeps changing. Across the stack, deployment environments change. Workloads shift, hardware generations evolve, toolchains update, and data distributions drift. This makes robust generalization and continual adaptation central. In LLM serving and cluster management, regression language models and trace-driven predictors highlight the promise of models that can adapt to new clusters with few examples, but they also raise questions about what invariants remain stable across deployments \citep{akhauri2025performance,clusterdata:Wilkes2020a}. In physical design, the community is explicitly measuring generalization across designs with suites like A2Perf, and exploring generative approaches like diffusion for broader transfer \citep{uchendu2024a2perf,lee_chip_2025}. In compiler and kernel optimization, brittleness across drivers, firmware, or topology suggests that learned systems must represent uncertainty and detect when they are out of distribution, rather than silently emitting plausible but fragile outputs \citep{shao2022metaschedule,ye2025flashinferefficientcustomizableattention}. The open question is how to design learning-based components that are robust to stack churn, can adapt continuously without catastrophic regressions, and expose when they should fall back to trusted baselines.

\paragraph{Open question 4: What does responsible human-model co-design look like for full-stack systems?}
Co-design across boundaries (C4) and role separation (P3) converge on a question that is as much organizational as it is technical. As GenAI systems become more agentic, the main product is not a single artifact but a workflow that mixes human intent, tool invocations, and model decisions. Across layers we see early versions of this. In software, IDE copilots and autonomous patching systems change how engineers debug and review code \citep{fan2023largelanguagemodelssoftware,singh2024coderesearcher}. In EDA, systems like LayoutCopilot emphasize interactive collaboration and explanation rather than full automation \citep{liu_layoutcopilot_2025}. In verification, vendor copilots sit atop existing engines and assist with triage, coverage closure, and specification review, suggesting that workflow integration is a primary path to impact \citep{Cadence2022Verisium,synopsys_copilot_generative_ai}. A cross-stack research agenda is to formalize provenance and accountability across these workflows. That includes tracking what the model proposed, what evidence supported it, what tools validated it, and which assumptions were made. It also includes designing interfaces that let humans specify constraints and intent in forms that verifiers can check, so that collaboration scales without turning into unstructured prompting.

\begin{table*}[t!]
\centering
\small
\setlength{\tabcolsep}{4pt}
\renewcommand{\arraystretch}{1.25}
\caption{The AI-for-Systems Readiness Matrix. Rows represent six core capabilities derived from the challenge--principle pairings in this survey, ordered by aggregate maturity. Columns represent evidence dimensions, ordered by pipeline progression. Ratings: \ding{108} established, \ding{119} emerging, \ding{109} early. See text for definitions and discussion.\vspace{5pt}}
\label{tab:readiness}
\hyphenpenalty=10000
\exhyphenpenalty=10000

\begin{tabular}{
>{\raggedright\arraybackslash}p{0.36\textwidth}
>{\centering\arraybackslash}p{0.105\textwidth}
>{\centering\arraybackslash}p{0.105\textwidth}
>{\centering\arraybackslash}p{0.105\textwidth}
>{\centering\arraybackslash}p{0.105\textwidth}
>{\centering\arraybackslash}p{0.105\textwidth}
}
\toprule

\multicolumn{1}{c}{\textbf{Required Capability}} &
\textbf{Methods} &
\textbf{Benchmarks} &
\textbf{Tools \& Infra.} &
\textbf{Deployment Evidence} &
\textbf{Cross-Domain Transfer} \\

\midrule

\textbf{1. Fast, structured feedback loop} \newline {\scriptsize Feedback Loop Crisis [C1] $\rightarrow$ Continuous Feedback [P2]} &
\ding{108} & \ding{119} & \ding{119} & \ding{119} & \ding{109} \\[6pt]

\textbf{2. Independent validation infrastructure} \newline {\scriptsize Trust \& Validation [C3] $\rightarrow$ Hybrid Approaches [P1]} &
\ding{119} & \ding{119} & \ding{119} & \ding{119} & \ding{109} \\[6pt]

\textbf{3. Adaptive policies with safety guarantees} \newline {\scriptsize Determinism to Dynamism [C5] $\rightarrow$ Hybrid Approaches [P1]} &
\ding{119} & \ding{109} & \ding{109} & \ding{119} & \ding{109} \\[6pt]

\textbf{4. Generator-verifier separation} \newline {\scriptsize Trust \& Validation [C3] $\rightarrow$ Separate Concerns by Role [P3]} &
\ding{119} & \ding{109} & \ding{109} & \ding{109} & \ding{109} \\[6pt]

\textbf{5. Captured and updatable domain knowledge} \newline {\scriptsize Tacit Knowledge [C2] $\rightarrow$ Build on Systems Knowledge [P5]} &
\ding{119} & \ding{109} & \ding{109} & \ding{109} & \ding{109} \\[6pt]

\textbf{6. Cross-layer optimization interfaces} \newline {\scriptsize Co-Design Across Boundaries [C4] $\rightarrow$ Match Problem Structure [P4]} &
\ding{119} & \ding{109} & \ding{109} & \ding{109} & \ding{109} \\

\bottomrule
\end{tabular}
\end{table*}

\subsection{The Systems Readiness Matrix}

The four questions above ask where the field needs to go. The Systems Readiness Matrix (Table~\ref{tab:readiness}) reframes the question as: what capabilities does any GenAI-for-systems pipeline need to demonstrate, and how mature are they?

The five columns are ordered by pipeline progression, from the earliest stage of maturity to the latest. \textit{Methods} asks whether proven algorithmic approaches exist for a capability; for feedback loops, this is established (compiler-in-the-loop, profiler-in-the-loop, and surrogate-guided methods are well-demonstrated), while for cross-layer interfaces it remains emerging (vertical co-design exists in a few industrial stacks but general methods are scarce). \textit{Benchmarks} asks whether standardized evaluations exist that specifically measure the capability, not just output quality; SWE-bench and KernelBench measure feedback loop effectiveness, but no benchmark yet measures how well tacit knowledge has been captured or how effectively a system transfers across layers. \textit{Tools and infrastructure} asks whether reusable frameworks, APIs, or validators are available that practitioners can adopt without rebuilding from scratch; formal verification engines and simulation harnesses provide this for validation, but no equivalent infrastructure exists for generator-verifier separation or cross-layer optimization. \textit{Deployment evidence} asks whether the capability has been demonstrated in production systems; feedback loops and adaptive policies have production instances (Google's ECO, learned memory allocators, LLM serving engines), while generator-verifier separation and cross-layer interfaces remain largely confined to research prototypes. \textit{Cross-domain transfer} asks whether an approach developed for one layer has been successfully applied to another; this is the column most directly tied to the systematic engineering argument of this survey, and it is early across the board.

The six rows represent capabilities that the survey identifies as necessary for reliable deployment. They are each derived from a challenge-principle pairing, and ordered from most mature to least mature.
\begin{itemize}
    \item A \textit{fast, structured feedback loop} (C1 $\rightarrow$ P2) is the ability to evaluate generated outputs and feed results back into the generation process, as in compiler-in-the-loop kernel optimization or simulator-in-the-loop RTL repair. 
    \item \textit{Independent validation infrastructure} (C3 $\rightarrow$ P1) is the availability of tools that can check generated outputs independently of the model that produced them, such as formal verification engines, simulation-based correctness testing, or robustness benchmarks with sanitizers. 
    \item \textit{Adaptive policies with safety guarantees} (C5 $\rightarrow$ P1) is the ability to deploy learned policies that adapt to changing conditions while maintaining predictability, as in hybrid congestion control or learned memory allocators with deterministic guard rails.
    \item \textit{Generator-verifier separation} (C3 $\rightarrow$ P3) is explicit architectural separation between the component that proposes and the component that judges, enabling modular trust and error localization, as in multi-agent RTL workflows or generate-then-check kernel pipelines.
    \item \textit{Captured and updatable domain knowledge} (C2 $\rightarrow$ P5) measures whether tacit expertise is encoded in datasets, retrieval systems, or tool interfaces rather than residing solely in practitioners, as in repo-scale benchmarks that expose implicit conventions or retrieval-augmented generation grounded in trusted IP. 
    \item \textit{Cross-layer optimization interfaces} (C4 $\rightarrow$ P4) is the ability to pass feedback and constraints across abstraction boundaries so that optimization at one layer accounts for effects at adjacent layers, as in joint model-system scheduling for LLM serving or placement-routing-timing co-optimization in chip design.
    
\end{itemize}

The table is designed to be read as a gradient: maturity decreases from top-left to bottom-right. Several patterns are visible. Methods (the leftmost column) are the most mature dimension across nearly all capabilities: the field has developed promising algorithmic approaches for feedback loops, validation, and adaptive policies. But methods alone are not enough. Benchmarks that specifically measure these capabilities, rather than just output quality, lag behind. Reusable tools and infrastructure are sparse outside of a few well-resourced areas. The approaches that work in one layer have rarely been formalized and ported to another, which is why the rightmost column remains the emptiest.

The least mature rows are also revealing. Captured and updatable domain knowledge (row 5) and cross-layer optimization interfaces (row 6) are early on nearly every dimension, confirming that these remain the hardest open problems in the field. Generator-verifier separation (row 4) has emerging methods but almost no standardized infrastructure, suggesting that the architectural pattern is understood but not yet engineered into reusable components.

The Systems Readiness Matrix makes the systematic engineering argument concrete. Rather than asking ``is generative AI ready for systems?'' as a single question, it decomposes readiness into the capabilities that actually matter and shows that progress is uneven. A research agenda organized around closing the gaps in this matrix, capability by capability and evidence dimension by evidence dimension, is what we mean by systematic engineering.

\section{Conclusion}
\label{sec:conclusion}

This survey set out to examine generative AI across the full computing stack, from software engineering and distributed systems through hardware architecture to RTL design, physical layout, and verification. The central finding is one of convergence. Despite the diversity of domains, tools, and research communities, the field keeps encountering the same five structural challenges and keeps arriving at the same five design principles as effective responses.

The challenge--principle map introduced in Section~\ref{sec:challenges_principles} makes these relationships explicit, and the body of the survey confirms the pattern. Across all eleven subsections spanning three layers, the same challenges and principles reappear with layer-specific instantiations but structurally similar dynamics. The feedback loop crisis (C1) drives investment in continuous feedback (P2) from agentic software loops to RTL tool-in-the-loop repair. Trust and validation (C3) gates deployment at every layer, consistently motivating hybrid verification (P1) and role separation between generators and checkers (P3). Co-design across boundaries (C4) surfaces wherever performance depends on decisions that span abstraction layers, and the responses that work match methods to the coupled structure of the problem (P4) rather than optimizing layers in isolation. The tacit knowledge problem (C2) and the shift from determinism to dynamism (C5) cut across all of these, shaping both what systems can learn and how they must adapt. The space of effective responses is far smaller than the space of problems, and that convergence can guide future work.

This survey is intentionally broad and emphasizes cross-layer synthesis over exhaustive coverage or detailed algorithmic comparison.\label{sec:limitation} We do not attempt to catalog all recent work, nor to rank models or methods within individual subdomains. Coverage within any single layer may appear selective, and many systems discussed remain research prototypes rather than fully production-hardened solutions. We focus primarily on technical and methodological considerations, leaving broader societal, economic, and environmental impacts largely out of scope. As models, benchmarks, and toolchains continue to evolve, some observations may require revision. We view this work as a snapshot and organizing framework intended to guide ongoing research rather than a definitive account of the field.

The cross-stack view is the main contribution. It lets us treat learning-based systems design as a unified problem of feedback, tacit knowledge, validation, boundary-spanning co-design, and dynamism, rather than as a collection of disconnected case studies. As argued in Section~\ref{sec:challenges_principles}, the field currently advances by ad hoc construction, with each community rediscovering lessons that others have already learned. The challenge--principle map, the open research questions, and the per-section cross-stack annotations are intended as a step toward the shared engineering methodology the field needs: common vocabularies, cross-layer benchmarks, and systematic design practices that let progress compound across communities. Generative AI changes systems not only by improving search and synthesis, but by making design loops adaptive and policy-driven across the stack. Making that transition reliable and systematic is the work ahead.

\section*{Acknowledgments}

This work originated from \href{https://harvard-edge.github.io/cs249r_fall2025/}{CS249r: Architecture 2.0, Agentic AI for Computer Systems Design}, a graduate seminar at Harvard University in Fall 2025. We are grateful to the guest speakers and industry practitioners who generously shared their expertise during the course, which directly informed the cross-stack analysis presented in this survey:
\textbf{Ofir Press} (Princeton) on code generation and LLM evaluation;
\textbf{Amir Yazdanbaksh} (Google DeepMind) on performance engineering and AI-driven systems design;
\textbf{Sasha Rush} (Cursor, Cornell Tech) on GPU kernel optimization and language models for code;
\textbf{Martin Maas} (Google DeepMind) on ML for runtime systems, operating systems, and computer architecture;
\textbf{Suvinay Subramanian} (Google DeepMind) on hardware accelerator performance modeling and design;
\textbf{Jenny Huang} (NVIDIA) on GPU architecture, accelerated computing, and hardware--software co-optimization;
\textbf{Milad Hashemi} (Google) on learned memory access patterns and ML for systems;
\textbf{Esha Choukse} (Microsoft Azure Research) on efficient AI workload scheduling and GenAI systems optimization;
\textbf{Mark Ren} (NVIDIA) on AI for chip design and GPU-accelerated EDA;
\textbf{Richard Ho} (OpenAI) on hardware--ML model co-optimization at scale;
and \textbf{Kartik Hegde} (ChipStack) on AI-assisted chip design and verification workflows.

We also thank the broader CS249r community for the weekly discussions, paper readings, and blog reflections that helped shape the challenges and principles distilled in this work. The course materials, weekly blog posts, and curated paper collection are available at \url{https://harvard-edge.github.io/cs249r_fall2025/}.

%%
%% Bibliography
%% NeurIPS uses plainnat for natbib compatibility
%%
\bibliographystyle{plainnat}
\bibliography{all_references}

\end{document}